\documentclass[preprint,12pt]{elsarticle}

\usepackage{amssymb}
\usepackage{amsfonts}
\usepackage{amsmath}
\usepackage{color}
\usepackage{booktabs}
\usepackage{mathbbol}
\usepackage{mathrsfs}
\usepackage{epsfig,bm,dcolumn}
\usepackage{graphicx}
\usepackage{geometry}
\usepackage{overpic}
\usepackage{slashed}
\newcommand{\upcite}[1]{\textsuperscript{\textsuperscript{\cite{#1}}}}
\usepackage{color}

\usepackage{algpseudocode}
\usepackage[ruled]{algorithm2e}
\journal{**}

\usepackage{caption}
\usepackage{float} 
\usepackage{subfig}
\usepackage{parskip}
\usepackage{multirow}
\usepackage[dvipsnames]{xcolor}

\begin{document}
\begin{sloppypar}
\begin{frontmatter}

\title{Approaching epidemiological dynamics of COVID-19 with physics-informed neural networks}
% \title{Studying Epidemiological Dynamics of COVID-19: Approaching Complicated Cases using PINNs}

\author[FIAS,IJC,GU,XD]{Shuai Han}
\author[FIAS,GU]{Lukas Stelz}
\author[FIAS,GU,GSI]{Horst Stoecker}
\author[FIAS,IJC]{Lingxiao Wang \corref{cor5}}
\author[FIAS,IJC]{Kai Zhou \corref{cor5}}
\cortext[cor5]{Corresponding author}
% \edd{zhou@fias.uni-frankfurt.de}

\address[FIAS]{Frankfurt Institute for Advanced Studies, Ruth-Moufang-Str. 1, 60438 Frankfurt am Main, Germany}
\address[IJC]{Xidian-FIAS International Joint Research Center, Ruth-Moufang-Str. 1, 60438 Frankfurt am Main, Germany}
% \address[GU]{Institute for Theoretical Physics, Goethe University, Max-von-Laue-Str. 1, 60438 Frankfurt am Main, Germany}
\address[GU]{Institut für Theoretische Physik, Goethe Universität Frankfurt, Max-von-Laue-Str. 1, 60438 Frankfurt am Main, Germany}
\address[GSI]{GSI Helmholtzzentrum für Schwerionenforschung GmbH, 64291 Darmstadt, Germany}
\address[XD]{Xidian University, 710071 Xi'an, China}

\begin{abstract}
A physics-informed neural network (PINN) embedded with the susceptible-infected-removed (SIR) model is devised to understand the temporal evolution dynamics of infectious diseases. Firstly, the effectiveness of this approach is demonstrated on synthetic data as generated from the numerical solution of the susceptible-asymptomatic-infected-recovered-dead (SAIRD) model. Then, the method is applied to COVID-19 data reported for Germany and shows that it can accurately identify and predict virus spread trends. The results indicate that an incomplete physics-informed model can approach more complicated dynamics efficiently. Thus, the present work demonstrates the high potential of using machine learning methods, e.g., PINNs, to study and predict epidemic dynamics in combination with compartmental models.

\end{abstract}


\end{frontmatter}

\section{Introduction}
\label{sec: introduction}
The coronavirus SARS-CoV-2 was discovered in Wuhan, China, in December 2019. The virus spread quickly around the world. It was declared a pandemic by the World Health Organization (WHO)\upcite{WHOcoronavirus2022} in March 2020. By January 2023, there had been 733 million confirmed cases of the resulting disease from COVID-19 and 6.69 million fatalities\upcite{WHOcoronavirus2022}. The spread of the virus and the impact of policy decisions on containing the disease has been studied in compartmental models\upcite{sedaghat2020predicting,siegenfeld2020models}. Non-pharmaceutical interventions, such as social distancing, were found to be effective\upcite{chinazzi2020effect,wilder2020isolation,zhang2020covid,banholzer2021estimating}. The role of vaccination has been explored in recent studies\upcite{thiel2021recommendations}.

The unknown numbers of infectious individuals have been a major challenge for obtaining precise real-time data on the spatiotemporal spread of COVID-19\upcite{cao2022covid}. 
Based on ambiguous reports and predicted cases\upcite{park2020time,dorn2023common}, governments had difficulties in implementing effective intervention policies, such as the allocation of detection resources\upcite{TANWAR2022108352}, the mobilization and delivery of protective and therapeutic materials\upcite{ertas2021role}, and the stringency of lockdown measures\upcite{kumar2020does,papadopoulos2020impact,barbarossa2020first,barbarossa2021germany}. Also the early shortage of detection supplies and the massive number of asymptomatic or mildly symptomatic cases also make forecasting difficult\upcite{peeri2020sars}. The alternative approach to the tracking of the spread of any infectious diseases is an estimate of parameters of epidemic models, such as the basic reproduction rate ($R_0$) and infection rate\upcite{xiang2021covid}. Studies presently explore various data-driven methods to infer those parameters from the available but limited data\upcite{cooper2020dynamic}.

Epidemiological models, such as the Susceptible-Infected-Recovered (SIR) model\upcite{kermack1927contribution}, have been helpful in understanding the spread of infectious diseases. The SIR model is one of the first used compartmental models. SIR divides the population into the three SIR compartments. Various models have been derived and developed, based on the SIR model, including the Susceptible-Exposed-Infected-Removed (SEIR)\upcite{aron1984seasonality,sun2020seir} and the Susceptible-Exposed-Asymptomatic-Infectious-Recovered (SEAIR) model\upcite{basnarkov2021seair}. Mathematical modeling based on numerical solutions of systems of Ordinary Differential Equations (ODEs) has also been used to study COVID-19 spread. These studies have provided valuable insights into the spread of the disease, including, more recently, disease prevalence curves with machine learning assistance\upcite{dandekar2020quantifying,jamshidi2020artificial,vaid2020deep}, the impact of asymptomatic infected individuals\upcite{stout2021silent}, the effectiveness of wearing masks\upcite{worby2020face}, and the effectiveness of prevention and control measures\upcite{oraby2021modeling,choi2021optimal,barbarossa2020modeling}.

Deep Learning (DL) models, such as recurrent neural networks (RNNs), have been used to analyze the patterns in COVID-19 time series data\upcite{hassan2020covid,wang2021machine}. Chimmula et al.\upcite{chimmula2020time} used RNNs and its variant long short-term memory (LSTM) for predicting COVID-19 prevalence trends in Canada and Italy, which show reasonable predictive capabilities. Zeroual et al.\upcite{zeroual2020deep} applied LSTM, bi-directional LSTM (BiLSTM), and gated recurrent units (GRUs) to different countries' COVID-19 data for data-driven simulations. Zhang et al.\upcite{qin2019data} used a residual neural network (ResNet) to account for external factors for model uncertainties, parameters, and for other factors which affect prediction accuracy for trend analysis of COVID-19. Chen et al.\upcite{chen2021generalized} showed that generalized ResNet can learn the structure of complex unknown dynamical systems. These predictions are more accurate than standard ResNet structures. However, these models require large sets of training data, while the current datasets for COVID-19 are relatively small. This leads to a lack of robustness of the models\upcite{nguyen2020artificial}. In addition, these DL models are able only to identify the dynamics of the virus based on available data; they might not be stable or accurate enough in predicting future trends\upcite{chen2021survey} due to the variability of the virus and the influence of external factors like weather\upcite{ganslmeier2021impact}. Frameworks that can accurately tackle the epidemic dynamics, which are governed by systems of ordinary differential equations (ODEs) or systems of partial differential equations (PDEs)\upcite{raissi2019physics}, should be developed to effectively handle the predicament induced by the limits of recorded data. It is crucial to incorporate the details of the necessarily known full domain knowledge, e.g., the laws governing the physical system, rigorously into the machine learning treatment used\upcite{wang2021physics}. Physics-Informed Neural Networks (PINNs)\upcite{karniadakis2021physics} were proposed to address this need. PINNs can introduce physical constraints into the training explicitly. The PINNs can make accurate predictions based on tiny datasets when Combined with epidemiological models\upcite{raissi2019physics}. They may even help to identify the underlying epidemiological dynamics\upcite{kharazmi2021identifiability}.

The present study introduces a SIR-dynamics-informed machine learning method to explore the approach towards complicated epidemiological dynamics, as often hidden in the study of the spread of COVID-19, by incorporating prior knowledge, in the form of the ODEs, from the SIR model into the loss functions of Deep Neural Networks (DNNs) as physical regularizer. This new method is first confronted with generated synthetic data of the ODE system, by using a SAIRD model to simulate different scenarios and test the effectiveness of the new approach. The method is then validated by reported COVID-19 data from Germany between March 1, 2021 and July 1, 2021. The data from the COVID-19 Data Repository at the Johns Hopkins University are used. Such simple, SIR-physics informed models, do can apparently approach more complicated dynamics efficiently.

This paper is organized into three sections: The methodology section details the general concepts of the SIR and SAIRD models and gives the definition of the parameters. The basics of the PINN framework, and how PINNs can be used to solve systems of ODEs and their optimization problems, is introduced in the second section. Here, the SIR-dynamics-informed neural networks, the definition and calculation of loss functions and physical residuals, which will be applied to estimate the dynamics of an infectious system are detailed. The third section contains the numerical synthetical data experiments and results for real, recorded data. It also covers the set-ups of the neural network, the pre-processing of synthetic and reported data, and the analysis of the results. The conclusions summarize the findings and yield recommendations for future research.

\section{Methodology}
\label{sec: Methodology}
This section introduces two mathematical epidemiological models used the present paper. Furthermore, physics-informed neural networks are described in conjunction with the epidemiological models and their optimization approaches.\par

% In this section we first introduce two mathematical epidemiological models used this study. Furthermore, we explain physics-informed neural networks in conjunction with the epidemiological models and their optimization approaches.\par

% This section first demonstrate two mathematical epidemiological models used in our study. Furthermore, we introduce phsyics-informed neural networks in conjunction with the epidemiological models and their optimisation approaches.\par

\subsection{Mathematical Epidemiology}
\label{sec:MathematicalEpidemiology}
Mathematical modelling of epidemiological dynamics is a popular area of research in applied mathematics. Conventionally, compartmental models are developed for modelling the dynamics of epidemics within a population.\par

% Mathematical modelling of epidemiological dynamics is a popular area of research in applied mathematics. Conventionally, compartmental models are developed for modelling the dynamics of epidemics within a population.\par

% Mathematical modelling of epidemiological dynamics is a popular area of research in applied mathematics, and compartmental modelling is the conventional way of modelling the dynamics of epidemics within a population.\par

\subsubsection{The SIR Model}
\label{sec: sir}
\begin{figure}[ht!]
	\centering
	\includegraphics[width=10cm]{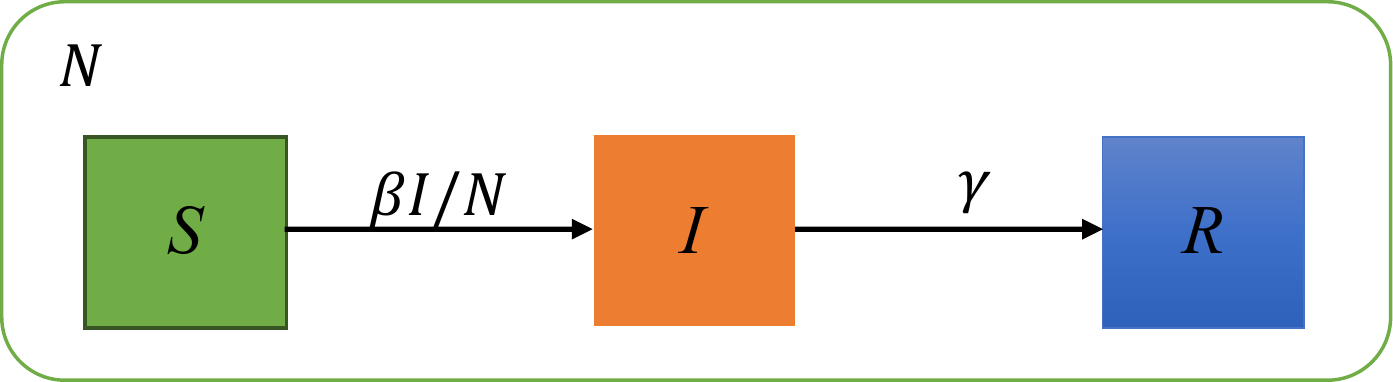}
	\caption{Schematic illustration of the interactions between the compartments in the SIR model.}
	\label{fig:sir}
\end{figure}

A plain, but yet powerful and well-known compartmental model is the original SIR model\upcite{peeri2020sars}. It assumes that the size of the population $N$ remains constant. $N$ is divided into three separate groups or compartments: susceptible (S), infectious (I) and removed (R). Individuals are transferred between compartments as shown in Figure~\ref{fig:sir} with certain rates, called transition rates, $\beta I/N$ and $\gamma$:\par

% A plain but yet powerful well known compartmental model is the SIR model. It is assumed that the size of the population $N$ remains constant and is divided into three separate groups or compartments: susceptible (S), infectious (I) and removed (R). Individuals are transferred between compartments as shown in Figure~\ref{fig:sir} with certain rates which are called transition rates. \par

% A plain but yet strong compartmental model is known as the SIR model. It is assumed that the size of the population remains constant, denoted as N. The population N is divided into three separate groups or compartments, respectively: susceptible (S), infectious (I) and removed or deaths (R) portions of the population. Individuals are transferred between compartments as shown in Figure \ref{fig:sir} at a certain rate which called the transition rate. \par
%%%%%%%%%%%%%%%%%%%%%%%%%%%%%%%%%%%%%%%%%%%%%%%%%%%%%%%%%%%%%%%%%%%%%%%%%%%%%%%%%%%%%%%%%%%
%%%%%%%%%%%%%%%%%%%%%%%%%%%%%%%%%%%%%%%%%%%%%%%%%%%%%%%%%%%%%%%%%%%%%%%%%%%%%%%%%%%%%%%%%%%
Epidemiological compartmental models treat all individuals in the same compartment as sharing identical features. Therefore each compartment is homogeneous. The SIR model is described by the following set of differential equations:

% In epidemiological compartmental models, all individuals in the same compartment are assumed to share identical features and therefore the whole compartment is homogeneous. The SIR model is described by the following set of differential equations:

%It occurs in the differential equations \ref{equ:sir}. In epidemiological compartmental models, It is assumed that all individuals in the same compartment share the identical features and therefore all groups are homogeneous. The SIR model can be described by the following differential equations:
%%%%%%%%%%%%%%%%%%%%%%%%%%%%%%%%%%%%%%%%%%%%%%%%%%%%%%%%%%%%%%%%%%%%%%%%%%%%%%%%%%%%%%%%%%%
\begin{equation}
    \begin{aligned}
    \frac{d S}{d t} &=-\frac{\beta I}{N} S\;, \\
    \frac{d I}{d t} &=\frac{\beta I}{N} S -\gamma I\;, \\
    \frac{d R}{d t} &=\gamma I\;.
    \end{aligned}
    \label{equ:sir}
\end{equation}
%%%%%%%%%%%%%%%%%%%%%%%%%%%%%%%%%%%%%%%%%%%%%%%%%%%%%%%%%%%%%%%%%%%%%%%%%%%%%%%%%%%%%%%%%%%

The parameter $\beta$ is the effective contact or transmission rate. It denotes the number of effective contacts made by one infectious and one susceptible individual leading to one infection per unit of time. The removal rate $\gamma$ indicates the fraction of infectious individuals who recover or die per unit of time. Thus, $\gamma$ can be calculated using 1/D, with D the average time duration that an infected individual can carry and transmit the virus. Equation~\eqref{equ:sir} is subject to the initial conditions, $ S\left(t_{0}\right) > 0, I\left(t_{0}\right) \geq 0, \text { and } R\left(t_{0}\right) \geq 0 $ at the initial time $ t_{0} $. The model assumes to conserve the total number of individuals, thus $S(t) + I(t) + R(t) = N$ holds at any time $ t $. In general, the time scale of the epidemic dynamics is assumed to be short as compared to the length of the life of individuals in the population: the effects of births and deaths on the population are simply ignored.\par

\subsubsection{The SAIRD Model}
\label{sec: saird}
\begin{figure}[ht!]
	\centering
	\includegraphics[width=10cm]{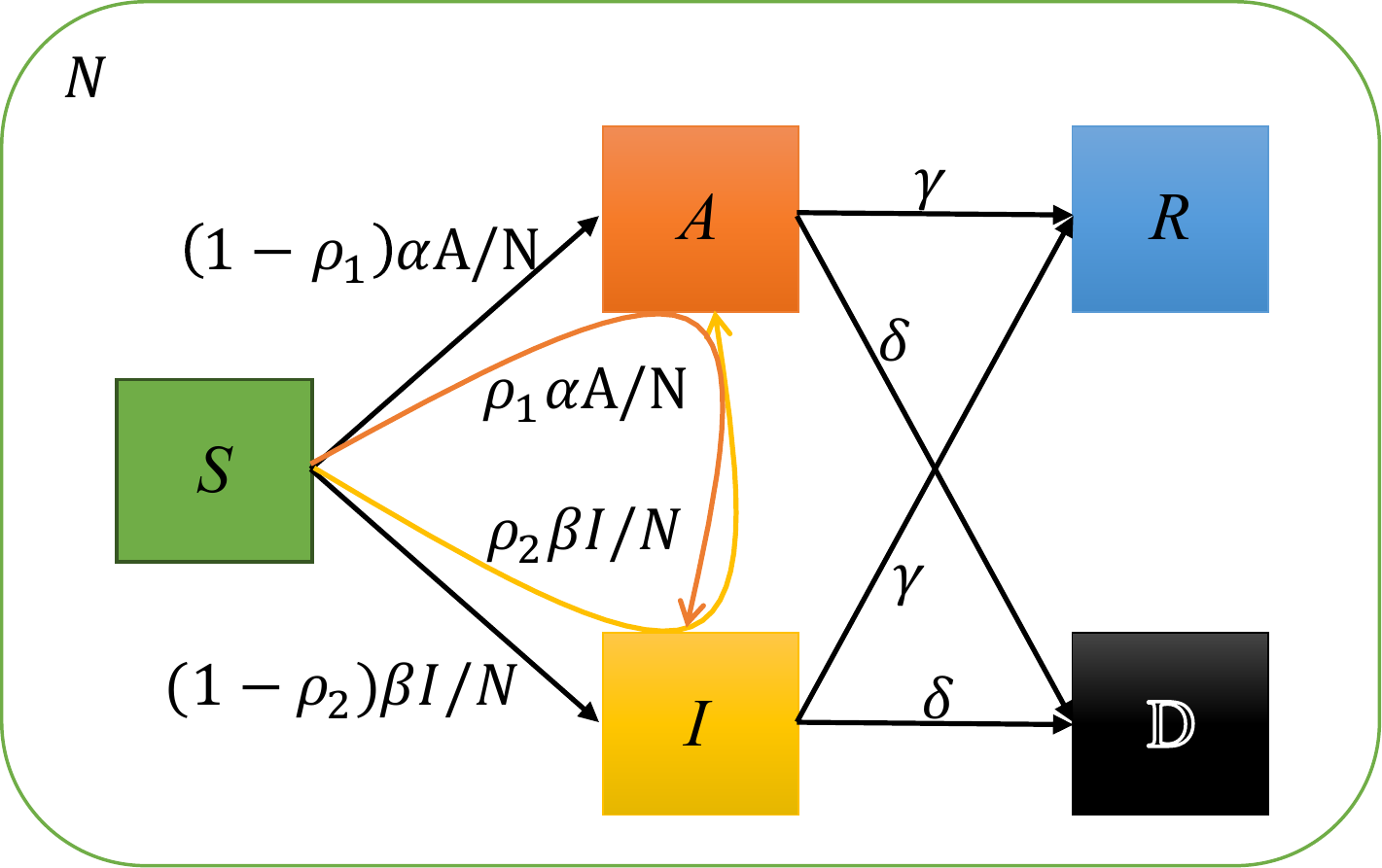}
	\caption{Diagram for the SAIRD model which is inspired from \upcite{angeli2022modeling} illustrating the interactions of compartment. }
	\label{fig:saird}
\end{figure}

The SAIRD model is an extension of the SIR model: two more compartments as shown in Figure~\ref{fig:saird} are introduced. The compartment $I$ is split into two. Here compartment $A$ represents the asymptomatic or unidentified, but infectious individuals, i.e., A is the number of individuals who, despite being infected, are either not identified or not detected. A consists of infected individuals, which are not from symptoms. The new compartment $I$ here contains those individuals which have been detected as infectious. The compartment $R$ is also split into the number of recovered, $R$, and the number of deceased, $D$, individuals.

% The SAIRD model extends the SIR model by introducing two more compartments as shown in Figure~\ref{fig:saird}. The $I$ compartment gets split into two. $ A $ represents the asymptomatic or unidentified infectious individuals, i.e. the number of individuals who, despite being infected, are not identified or detected. This occurs mainly for those infected individuals who are not suffering any symptoms. The new $I$ compartment contains only the detected infectious individuals. Similarly, the $R$ compartment is split into recovered $R$ and deceased $ D $, i.e. the number of individuals that have died.

% The SAIRD model extends the SIR model by introducing two new compartments as additional compartments. As shown in Figure \ref{fig:saird}, $ A $ represents the compartment of asymptomatic or unidentified individuals, i.e. the number of individuals who, despite being infected, are not identified or detected. This occurs mainly in infected individuals who are not suffering any symptoms. $ D $ stands for the number of individuals that have died.
%%%%%%%%%%%%%%%%%%%%%%%%%%%%%%%%%%%%%%%%%%%%%%%%%%%%%%%%%%%%%%%%%%%%%%%%%%%%%%%%%%%%%%%%%%%%
%%%%%%%%%%%%%%%%%%%%%%%%%%%%%%%%%%%%%%%%%%%%%%%%%%%%%%%%%%%%%%%%%%%%%%%%%%%%%%%%%%%%%%%%%%%%
The SAIRD model is given by the following set of differential equations:
%%%%%%%%%%%%%%%%%%%%%%%%%%%%%%%%%%%%%%%%%%%%%%%%%%%%%%%%%%%%%%%%%%%%%%%%%%%%%%%%%%%%%%%%%%%%
\begin{equation}
    \begin{array}{l}
    \frac{d S}{d t} \vspace{1ex}=-\beta I \frac{S}{N}- {\alpha} A \frac{S}{N}\;, \\
    \frac{d A}{d t}  \vspace{1ex} =(1-\rho_{1}) \alpha A \frac{S}{N}+ \rho_{2} \beta I \frac{S}{N}-\gamma A-\delta A\;, \\
    \frac{d I}{d t} \vspace{1ex} =(1-\rho_{2}) \beta I \frac{S}{N} + \rho_{1}\alpha A \frac{S}{N} -\gamma I-\delta I\;, \\
    \frac{d R}{d t} \vspace{1ex} =\gamma I+\gamma A\;, \\
    \frac{d D}{d t} \vspace{1ex} =\delta I+\delta A\;.
    \end{array}
    \label{equ:saird}
\end{equation}
%%%%%%%%%%%%%%%%%%%%%%%%%%%%%%%%%%%%%%%%%%%%%%%%%%%%%%%%%%%%%%%%%%%%%%%%%%%%%%%%%%%%%%%%%%%%
The transition rates and flow between the compartments are shown in Figure~\ref{fig:saird}. The infections happen at a rate of $\alpha$ and $\beta$ is due to contact of a susceptible individual with an asymptomatic or symptomatic infectious individual, respectively. The probability of a susceptible individual to become a symptomatically infected individual, through contact with an asymptomatic individual, is $\rho_1$. The probability of a susceptible individual to become an asymptomatically infected individual, through contact with a symptomatically infected individual, is $\rho_2$. Infected individuals recover at a rate of $\gamma$. They will pass away at a rate of $\delta$, independently of their symptoms. This system~\eqref{equ:saird} is subject to the initial conditions $S\left(t_{0}\right) > 0, A\left(t_{0}\right) \geq 0, I\left(t_{0}\right) \geq 0, R\left(t_{0}\right) \geq 0 \text { and } D\left(t_{0}\right) \geq 0$, with an initial time $ t_{0} $. The SAIRD model assumes, just as, that the total population, including the deceased, remains constant, $N$. Hence, this model satisfies $S(t) + A(t) +I(t) + R(t) + D(t) = N$, at any time $t$. \par

\subsection{Physics-Informed Neural Network}
\label{sec: pinn}
\begin{figure}[ht!]
	\centering
	\includegraphics[width=12cm]{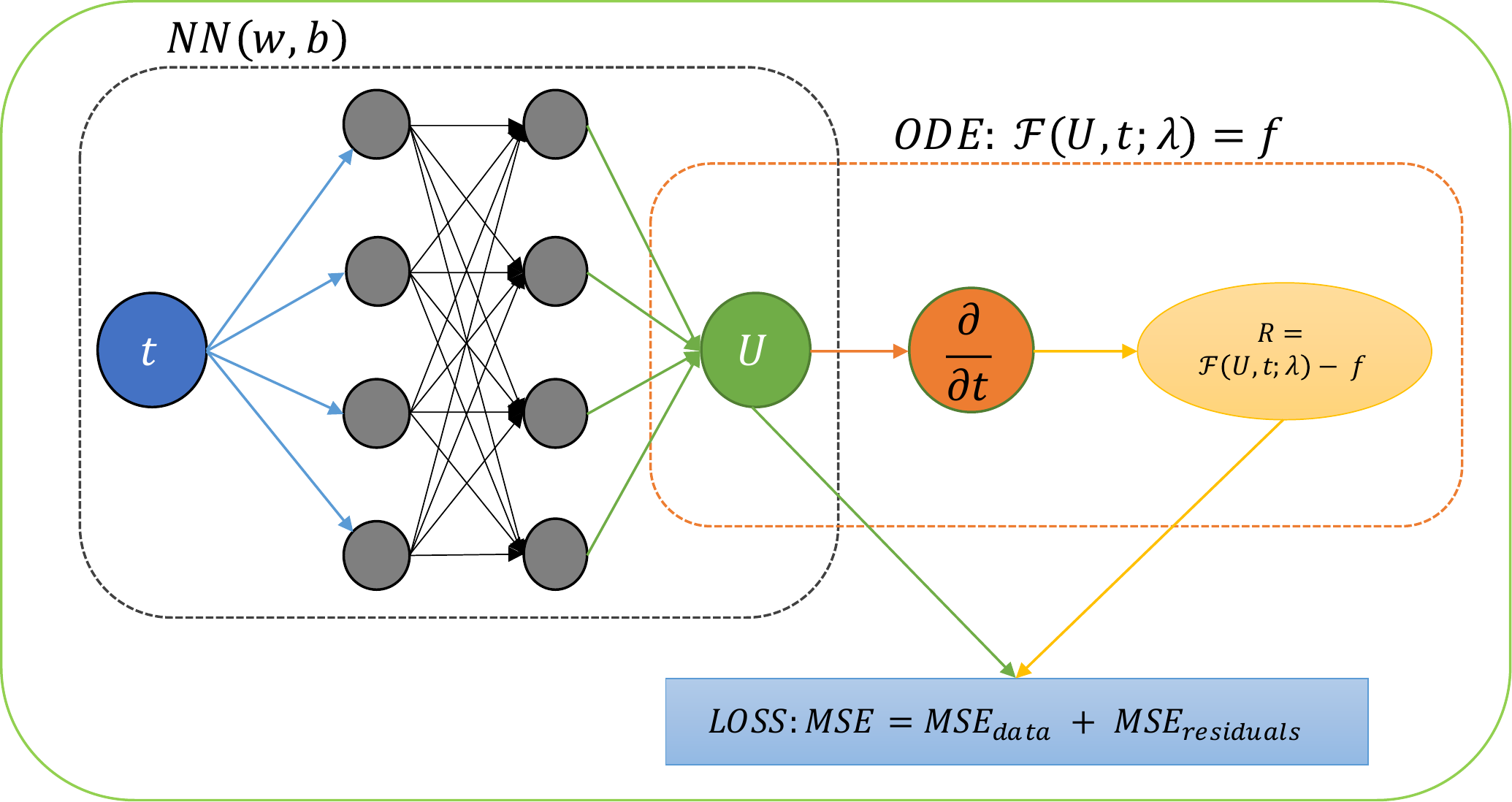}
	\caption{A schematic diagram of physics-informed neural networks (PINNs). The black dashed-line block is a common neural network that takes a time $t$ as input and the output is $U$, $\beta$ and $\gamma$ are weights and biases, respectively. The orange dashed-line block stands for the calculation of residual loss. The loss function consists of a mismatch of boundaries and initial conditions for the observed data (data loss). The residuals of the ODE is a set of random points in the spatial-temporal domain (residuals). The parameters of the PINNs can be optimized by minimising the loss $ MSE = MSE_{{data }}+MSE_{{residuals}}$.}
	\label{fig:pinns}
\end{figure}

The basic idea of PINNs is to integrate any a-priori knowledge of the system into the learning process of the deep neural network. This knowledge, e.g., about basic physical laws or domain know-how, is usually given in the form of ordinary- or partial differential equations (ODEs/PDEs), and is incorporated in PINNs by the loss functions. The training is performed to optimize the network weights and biases, and also to optimize the model parameters (e.g., those inside the physical laws). The loss term consists of two terms, the data loss and the residual loss, representing the regularization from the physics prior, like the generally obeyed differential equations. Figure~\ref{fig:pinns} illustrates the training process for an ODE-dynamics-informed neural network.

% The basic idea of PINNs is to integrate a-priori knowledge of the system into the learning process of a deep neural network. This knowledge about basic physical laws or domain know-how is usually given in the form of ordinary or partial differential equations (ODEs/PDEs), and in PINNs is incorporate into the loss function. The training is performed to optimize the network weights and biases, and also the model parameters (e.g. inside the physical laws). The loss term consists of two terms, the data loss and the residual loss representing regularization from the physics prior like the generally obeyed differential equations. Figure~\ref{fig:pinns} illustrates the training process for an ODE-dynamics-informed neural network.

% The fundamental idea of PINNs, as we can see in Figure \ref{fig:pinns} (Example for ODEs), is to integrate a priori knowledge in the form of basic physical laws or domain know-how modelled by ordinary or partial differential equations (ODEs/PDEs) into deep learning models. Especially, this is achieved by training a neural network with respect to their input variables and model parameters. Then, over and above data loss, the residuals of the differential equation are minimized in a least squares sense as part of the loss function.
%%%%%%%%%%%%%%%%%%%%%%%%%%%%%%%%%%%%%%%%%%%%%%%%%%%%%%%%%%%%%%%%%%%%%%%%%%%%%%%%%%%%%%%%%%
%%%%%%%%%%%%%%%%%%%%%%%%%%%%%%%%%%%%%%%%%%%%%%%%%%%%%%%%%%%%%%%%%%%%%%%%%%%%%%%%%%%%%%%%%%
Figure~\ref{fig:pinns} shows how a PINN can be used to tackle a solution of the system of ODEs by training the neural network using the combined losses. The system of first-order ordinary differential equations of the general form is usually written as:

% As shown in Figure~\ref{fig:pinns}, a PINN can be used to tackle a system of ODEs by training the neural network using the combined loss. A system of first-order ordinary differential equations of the general form is usually written as follows:
%%%%%%%%%%%%%%%%%%%%%%%%%%%%%%%%%%%%%%%%%%%%%%%%%%%%%%%%%%%%%%%%%%%%%%%%%%%%%%%%%%%%%%%%%%
\begin{equation}
    \begin{aligned}
        \frac{\partial U}{\partial t} (t)+F(U(t);\lambda)=0, \quad t\in [t_{0},T]
    \end{aligned}
    \label{equ:first-order}
\end{equation}
%%%%%%%%%%%%%%%%%%%%%%%%%%%%%%%%%%%%%%%%%%%%%%%%%%%%%%%%%%%%%%%%%%%%%%%%%%%%%%%%%%%%%%%%%%%
with 
%%%%%%%%%%%%%%%%%%%%%%%%%%%%%%%%%%%%%%%%%%%%%%%%%%%%%%%%%%%%%%%%%%%%%%%%%%%%%%%%%%%%%%%%%%%
\begin{equation}
    U(t) = [u^{1}(t),..., u^{n}(t)], \quad
    F(U) = [f^{1}(U),...,f^{n}(U)]
\end{equation}
%%%%%%%%%%%%%%%%%%%%%%%%%%%%%%%%%%%%%%%%%%%%%%%%%%%%%%%%%%%%%%%%%%%%%%%%%%%%%%%%%%%%%%%%%%%
where $ u^{i} \in \mathbb{R} $ and  $ f^{i} : \mathbb{R} \rightarrow \mathbb{R}, i=1,...,n $. $t_{0}$ and $T$ are the initial and final time, respectively; $F$ is the function and $ U $ the solution. The $ \lambda \in \mathbb{R}^{k} $ are the unknown parameters of the system \eqref{equ:first-order}. $U_{s}$ are the observed data at times $t_{1},...,t_{m}$, which determine $\lambda$. The data loss is, naturally,

% where $ u^{i} \in \mathbb{R} $ and  $ f^{i} : \mathbb{R} \rightarrow \mathbb{R}, i=1,...,n $. $ t_{0} $ and $ T $ are the initial and final time, respectively. Moreover, $ F $ is the function and $ U $ the solution, $ \lambda \in \mathbb{R}^{k} $ are the unknown parameters for the system \eqref{equ:first-order}. Denote $ U_{s} $ to be the observed data at time points $t_{1},...,t_{m}$, in determining $\lambda$ the data loss part is naturally
%%%%%%%%%%%%%%%%%%%%%%%%%%%%%%%%%%%%%%%%%%%%%%%%%%%%%%%%%%%%%%%%%%%%%%%%%%%%%%%%%%%%%%%%%%%
\begin{equation}
    \begin{aligned}
        \mathcal{L}_{data}= \sum_{s=1}^{m}\left \| U(t_{s}) - U_{s} \right \|^2 \;. 
    \end{aligned}
    \label{equ:diff_IO}
\end{equation}
%%%%%%%%%%%%%%%%%%%%%%%%%%%%%%%%%%%%%%%%%%%%%%%%%%%%%%%%%%%%%%%%%%%%%%%%%%%%%%%%%%%%%%%%%%%
For a conventional fitting, other than with PINNs, one identifies the optimum vector of the model parameters $\lambda$ by minimizing eq.\eqref{equ:diff_IO}. Thus, a solution $U(t)$ is obtained which is best suitable for the observed data, in the sense of the least squares deviations. The PINNs are based on a general neural network, as shown in Figure~\ref{fig:pinns} (black dashed frame). Its form can be represented as follows:

% For a conventional fitting other than PINNs, one just identifies the optimal vector of the model parameters $\lambda$ by minimizing \eqref{equ:diff_IO}, thus obtaining a solution $U(t)$ that is the best suitable for the observed data in the sense of least squares. The PINNs is based on a general neural network in Figure~\ref{fig:pinns} (black dashed frame), the form of which can be represented as follows:

% It can be understood simply, we identify the optimal vector of trainable parameter $\lambda$ by minimizing \eqref{equ:diff_IO}, thus obtaining a solution $U$ that is the best suitable for our observed data in the sense of least squares.
% The PINNs is based on a general neural network in Figure~\ref{fig:pinns} (Black dash line frame), the form of which can be represented as follows:
%%%%%%%%%%%%%%%%%%%%%%%%%%%%%%%%%%%%%%%%%%%%%%%%%%%%%%%%%%%%%%%%%%%%%%%%%%%%%%%%%%%%%%%%%%%
\begin{equation}
    \begin{aligned}
        {N\!N}^{\omega , b}(t): \mathbb{R} \rightarrow \mathbb{R}^{n}
    \end{aligned}
    \label{equ:nn}
\end{equation}
%%%%%%%%%%%%%%%%%%%%%%%%%%%%%%%%%%%%%%%%%%%%%%%%%%%%%%%%%%%%%%%%%%%%%%%%%%%%%%%%%%%%%%%%%%%
which approximates the solution
%%%%%%%%%%%%%%%%%%%%%%%%%%%%%%%%%%%%%%%%%%%%%%%%%%%%%%%%%%%%%%%%%%%%%%%%%%%%%%%%%%%%%%%%%%%
\begin{equation}
    \begin{aligned}
        U(t): \mathbb{R} \rightarrow \mathbb{R}^{n}
    \end{aligned}
\end{equation}
%%%%%%%%%%%%%%%%%%%%%%%%%%%%%%%%%%%%%%%%%%%%%%%%%%%%%%%%%%%%%%%%%%%%%%%%%%%%%%%%%%%%%%%%%%%
of the system of first-order ODEs. The weights $\omega$ and biases $b$ of $ {N\!N}^{\omega, b} $ are trainable parameters of the neural network. For the purpose of solving ODEs like \eqref{equ:first-order} with neural networks \eqref{equ:nn}, the weights $\omega$ and biases $b$ can be optimized and that the neural network \eqref{equ:nn} offers the best fits of the observed data $ U_{s}, s=1, ..., m $, in the sense of the least squares differences,

% of the system of first-order ODEs. The weights $\omega$ and biases $b$ of $ {N\!N}^{\omega, b} $ are trainable parameters of the neural network. For the purpose of solving ODEs like \eqref{equ:first-order} with neural networks \eqref{equ:nn}, one can optimize the weights $\omega$ and biases $b$ so that the neural network \eqref{equ:nn} best fits the observed data $ U_{s}, s=1, ..., m $, in the sense of least squares, %Then, we would like to optimize

% of the system of first-order ODEs. The weights $w$ and biases $b$ of $ {N\!N}^{\omega, b} $ can be used to determine the neural network. For purpose of solving ODEs \eqref{equ:first-order} by neural network \eqref{equ:nn}, we can optimize the weights $\omega$ and biases $b$ so that neural network \eqref{equ:nn} can fit the observed data $ \dot{U_{s}}, s=1, ..., m $, in the sense of least squares. Then, we would like to optimize
%%%%%%%%%%%%%%%%%%%%%%%%%%%%%%%%%%%%%%%%%%%%%%%%%%%%%%%%%%%%%%%%%%%%%%%%%%%%%%%%%%%%%%%%%%%
\begin{equation}
   \begin{aligned}
       %\omega,b = 
       \operatorname*{arg\,min}_{\omega, b} (\text{MSE}_{U}^{\omega, b})
   \end{aligned}
   \label{equ:minimize_mse_IO}
\end{equation}
%%%%%%%%%%%%%%%%%%%%%%%%%%%%%%%%%%%%%%%%%%%%%%%%%%%%%%%%%%%%%%%%%%%%%%%%%%%%%%%%%%%%%%%%%%%
Thus the loss function with respect to the observed data is
%%%%%%%%%%%%%%%%%%%%%%%%%%%%%%%%%%%%%%%%%%%%%%%%%%%%%%%%%%%%%%%%%%%%%%%%%%%%%%%%%%%%%%%%%%%
\begin{equation}
    \begin{aligned}
        \text{MSE}_{U}^{\omega, b} := \frac{1}{m} \sum_{s=1}^{m}\left\| {N\!N}^{\omega, b}\left(t_{s}\right)- {U}_{s}\right\|^{2}.
    \end{aligned}
    \label{equ:mse_IO}
\end{equation}
%%%%%%%%%%%%%%%%%%%%%%%%%%%%%%%%%%%%%%%%%%%%%%%%%%%%%%%%%%%%%%%%%%%%
The residual loss of the set of ODEs in \eqref{equ:first-order}, which can be expressed as

% We further include the residual loss of the set of ODEs in \eqref{equ:first-order}, which ban be expressed as
%%%%%%%%%%%%%%%%%%%%%%%%%%%%%%%%%%%%%%%%%%%%%%%%%%%%%%%%%%%%%%%%%%%%%%%%%%%%%%%%%%%%%%%%%%%
\begin{equation}
    \begin{aligned}
        \mathcal{F}\left({N N}^{\omega, b}, t;\lambda \right)=\frac{\partial {N N}^{\omega, b}}{\partial t}(t)+F\left({N N}^{\omega , b}(t)\right),
    \end{aligned}
    \label{equ:mse_residual}
\end{equation}
%%%%%%%%%%%%%%%%%%%%%%%%%%%%%%%%%%%%%%%%%%%%%%%%%%%%%%%%%%%%%%%%%%%%%%%%%%%%%%%%%%%%%%%%%%%
allows to extend the ${NN}^{\omega, b}$ to a PINN by putting the residual term \eqref{equ:mse_residual} into the loss function \eqref{equ:mse_IO}. The automatic differentiation technique of the neural networks can be used to compute the derivatives ($\frac{\partial {N N}^{\omega, b}}{\partial t}$) of the output of the network with respect to the input  (see Figure~\ref{fig:pinns}). Thus, letting $\mathcal{F}\left({NN}^{\omega, b}, t;\lambda \right)=0, \forall t \in[t_{0},T]$ is equivalent to forcing a neural network \eqref{equ:nn} to fulfill the ODE dynamics \eqref{equ:first-order}. \par

In other words, the standard neural network can be turned into a PINN by adding a mean squared residual error $\text{MSE}_{\mathcal{F}}^{\omega, b,\lambda}$ to the loss function \eqref{equ:minimize_mse_IO}. Thus, PINNs can be trained to identify the optimum neural network parameters, $\omega$ and $b$, as well as the parameters $\lambda$ for the ODEs. In that way, the following ODE-dynamics-regularized optimization is solved:

% this allows us to extend the ${NN}^{\omega, b}$ to a PINN by putting the residual term \eqref{equ:mse_residual} into the loss function \eqref{equ:mse_IO}. We can use the automatic differentiation technique of neural networks to compute the derivatives ($\frac{\partial {N N}^{\omega, b}}{\partial t}$) of the output of the network with respect to the input  (like Figure~\ref{fig:pinns}). Thus, letting $\mathcal{F}\left({NN}^{\omega, b}, t;\lambda \right)=0, \forall t \in[t_{0},T]$ is equivalent to forcing a neural network \eqref{equ:nn} to fulfill the ODE dynamics \eqref{equ:first-order}. \par

% In other words, we turned the standard neural network into a PINN by adding mean squared residual error $\text{MSE}_{\mathcal{F}}^{\omega, b,\lambda}$ to the loss function \eqref{equ:minimize_mse_IO}. Thus, we can train PINNs to identify the optimal neural network parameters $\omega$ and $b$ as well as the parameters $\lambda$ for ODEs, that is to solve the following ODE-dynamics-regularized optimization,
%%%%%%%%%%%%%%%%%%%%%%%%%%%%%%%%%%%%%%%%%%%%%%%%%%%%%%%%%%%%%%%%%%%%%%%%%%%%%%%%%%%%%%%%%%%
\begin{equation}
    \begin{aligned}
        %\omega,b,\lambda = 
        \operatorname*{arg\,min}_{\omega, b,\lambda } (\text{MSE}_{U}^{\omega, b}+\text{MSE}_{\mathcal{F}}^{\omega, b,\lambda}).
    \end{aligned}
\end{equation}
%%%%%%%%%%%%%%%%%%%%%%%%%%%%%%%%%%%%%%%%%%%%%%%%%%%%%%%%%%%%%%%%%%%%%%%%%%%%%%%%%%%%%%%%%%%

\subsection{SIR Model-Informed Machine Learning}
\label{sec: sir-iml}
This section explicitly introduces how to incorporate the SIR model into the PINN formalism, as a prior information. The dynamic parameters of the SIR model are estimated. This includes, both, the definition and the calculation of the data loss function and of the physical residuals. 

% In this section, we explicitly introduce how to incorporate the SIR model into the PINN formalism as a prior information, and estimate the dynamic parameters of the SIR model. This includes the definition and calculation of the data loss function and physical residuals. 

% In this section, we first explicitly introduce how to put the SIR model into the PINN formalism as a physical law, and estimate the dynamic parameters of the SIR model, as well as the definition and calculation of the loss function and physical residuals. 
%%%%%%%%%%%%%%%%%%%%%%%%%%%%%%%%%%%%%%%%%%%%%%%%%%%%%%%%%%%%%%%%%%%%%%%%%%%%%%%%%%%%%%%%%%%
\begin{figure}[ht!]
	\centering
	\includegraphics[width=15cm]{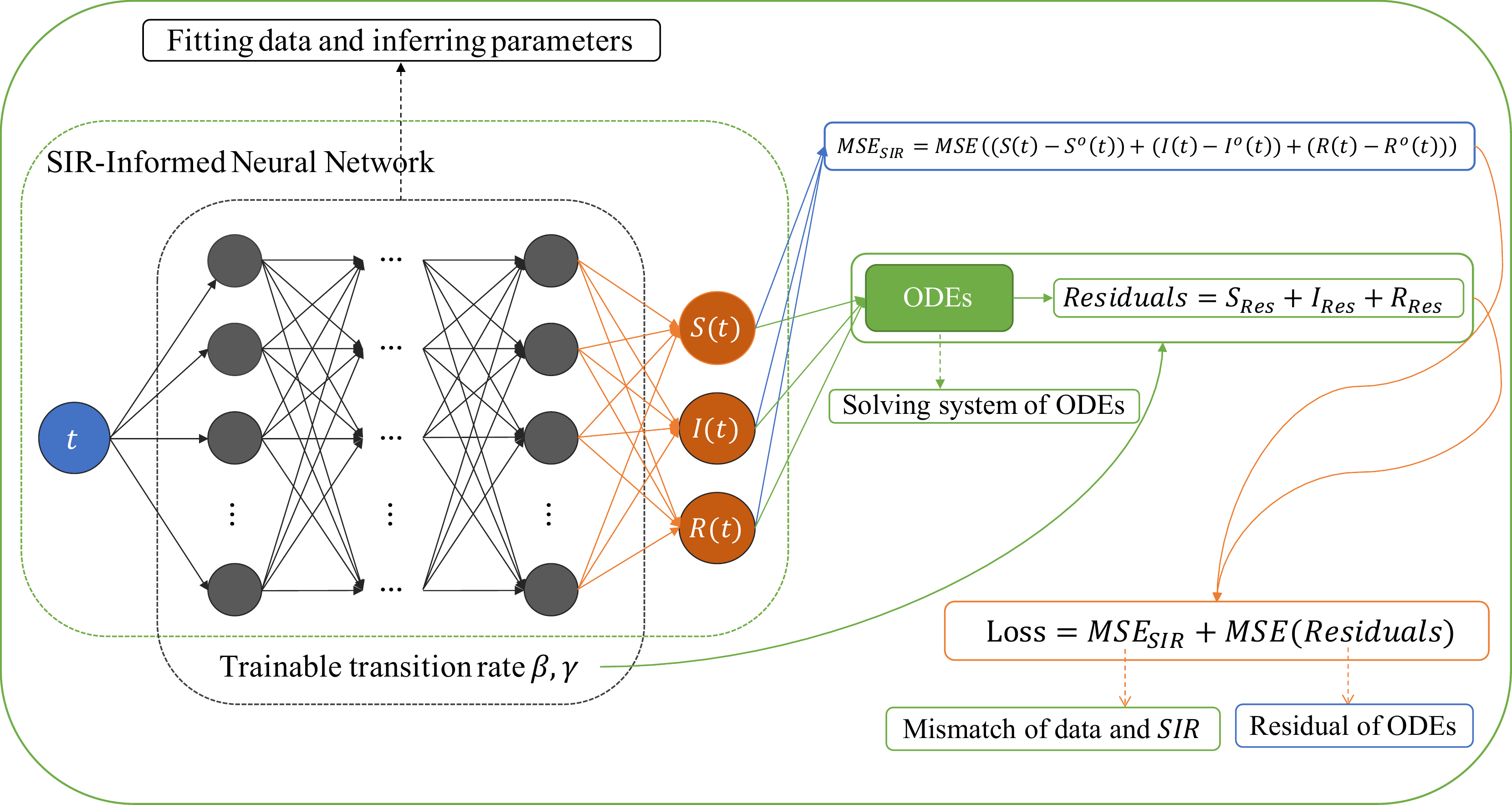}
	\caption{Schematic diagram of the SIR-dynamics informed neural network. The black-dashed frame represents the dense neural network used here. The green-dashed frame, on the other hand, represents the SIR-informed neural network, which takes time $t$ as input and outputs the susceptible-$(S)$, infected-$(I)$ and removed $(R)$ populations. The box labeled ‘ODEs’ represents the computation of the residual, with respect to the SIR model. The label ‘Loss’ is comprised of two parts: the mismatch between the available data and the network output, on one hand and the physical residual, on the other hand. The NN fits simultaneously both, the data and infers the dynamic parameters $\beta$ and $\gamma$, by satisfying the ODE dynamics, by minimizing loss function.}
	\label{fig:sir_informed_pinns}
\end{figure}

% Schematic diagram of the SIR-dynamics-informed neural network. Black dashed frame represents the dense neural network used. The green dashed frame represents the SIR-informed neural network, which takes an input of time $t$ and outputs susceptible $(S)$, infected $(I)$ and removed $(R)$. The ‘ODEs’ box represents computation of the residual with respect to SIR model. The ‘Loss’ comprises of two parts: the mismatch between available data and the network output, and the physical residual. By minimizing this loss function, the NN simultaneously fits the data and infers the dynamic parameters $\beta$ and $\gamma$ by satisfying ODE dynamics.
%%%%%%%%%%%%%%%%%%%%%%%%%%%%%%%%%%%%%%%%%%%%%%%%%%%%%%%%%%%%%%%%%%%%%%%%%%%%%%%%%%%%%%%%%%%
\subsection{Architecture}
\label{sec: architeture}
Figure~\ref{fig:sir_informed_pinns} utilize, for the SIR model, a fully connected neural network (marked by the black-dashed frame), to evaluate $(S(t), I(t), R(t))^\top$ as defined in \eqref{equ:nn}. Here, $S(t)$, $I(t)$ and $R(t)$ obey the SIR model at a given input time $t$. The residual term of the ODEs can be minimized for the SIR model as defined in \eqref{equ:mse_residual} to enforce eq.\eqref{equ:nn}. Thus, here

% As shown in Figure~\ref{fig:sir_informed_pinns}, for the SIR model, we utilized a fully connected neural network (marked within the black dashed frame), to evaluate $(S(t),I(t),R(t))^\top$ defined in \eqref{equ:nn}, where $S(t)$, $I(t)$ and $R(t)$ are supposed to obey the SIR model at a given input $t$. We can minimize the residual term of the ODEs for SIR model defined in \eqref{equ:mse_residual} to enforce \eqref{equ:nn}, here thus
%%%%%%%%%%%%%%%%%%%%%%%%%%%%%%%%%%%%%%%%%%%%%%%%%%%%%%%%%%%%%%%%%%%%%%%%%%%%%%%%%%%%%%%%%%%
\begin{equation}
       \mathcal{F}\left({N N}^{\omega,b}, t; \beta,\gamma\right)=
        \begin{bmatrix}
            \frac{d S(t)}{d t}+\frac{\beta S(t) I(t)}{N} \\ 
            \frac{d I(t)}{d t}-\frac{\beta S(t) I(t)}{N}+\gamma I(t) \\
            \frac{d R(t)}{d t}-\gamma I(t) 
        \end{bmatrix}.
\end{equation}
%%%%%%%%%%%%%%%%%%%%%%%%%%%%%%%%%%%%%%%%%%%%%%%%%%%%%%%%%%%%%%%%%%%%%%%%%%%%%%%%%%%%%%%%%%%
Hence, the mean residual squared error of the present work is
%%%%%%%%%%%%%%%%%%%%%%%%%%%%%%%%%%%%%%%%%%%%%%%%%%%%%%%%%%%%%%%%%%%%%%%%%%%%%%%%%%%%%%%%%%%
\begin{equation}
    \begin{aligned}
    \text{MSE}_{SIR} &=\text{MSE}_{S_{residual}}+\text{MSE}_{I_{residual}}+\text{MSE}_{R_{residual}}
    \end{aligned}
\end{equation}
%%%%%%%%%%%%%%%%%%%%%%%%%%%%%%%%%%%%%%%%%%%%%%%%%%%%%%%%%%%%%%%%%%%%%%%%%%%%%%%%%%%%%%%%%%%
with
%%%%%%%%%%%%%%%%%%%%%%%%%%%%%%%%%%%%%%%%%%%%%%%%%%%%%%%%%%%%%%%%%%%%%%%%%%%%%%%%%%%%%%%%%%%
\begin{equation}
    \begin{aligned}
        \text{MSE}_{S_{residual}} &=\frac{1}{q} \sum_{i=1}^{q} \left|\frac{d {S}(t_{i})}{d t_{i}}+\frac{\beta {S}(t_{i}) {I}(t_{i})} {N} \right|^{2}, \\
        \text{MSE}_{I_{residual}} &=\frac{1}{q} \sum_{i=1}^{q} \left|\frac{d {I}(t_{i})}{d t_{i}}-\frac{\beta {S}(t_{i}) {I}(t_{i})} {N} + \gamma I(t_{i})\right|^{2}, \\
        \text{MSE}_{R_{residual}} &=\frac{1}{q} \sum_{i=1}^{q} \left|\frac{d {R}(t_{i})}{d t_{i}}-\gamma I(t_{i}) \right|^{2}.
    \end{aligned}
\end{equation}
%%%%%%%%%%%%%%%%%%%%%%%%%%%%%%%%%%%%%%%%%%%%%%%%%%%%%%%%%%%%%%%%%%%%%%%%%%%%%%%%%%%%%%%%%%
Here $q$ is the total number of discrete time points. Note that, the discrete time points are chosen to be consistent with the observed time step, chosen in units of one natural day. In other words, the time step between two time points is $\Delta t$=1. \par

Simultaneously, the mean squared error of the data is, for the SIR model, 
%%%%%%%%%%%%%%%%%%%%%%%%%%%%%%%%%%%%%%%%%%%%%%%%%%%%%%%%%%%%%%%%%%%%%%%%%%%%%%%%%%%%%%%%%%
\begin{equation}
    \text{MSE}_{data} =\text{MSE}_{S_{data}}+\text{MSE}_{I_{data}}+\text{MSE}_{R_{data}}.
\end{equation}
%%%%%%%%%%%%%%%%%%%%%%%%%%%%%%%%%%%%%%%%%%%%%%%%%%%%%%%%%%%%%%%%%%%%%%%%%%%%%%%%%%%%%%%%%%
Here 
%%%%%%%%%%%%%%%%%%%%%%%%%%%%%%%%%%%%%%%%%%%%%%%%%%%%%%%%%%%%%%%%%%%%%%%%%%%%%%%%%%%%%%%%%%
\begin{equation}
    \begin{aligned}
        \text{MSE}_{S_{data}} &=\frac{1}{s} \sum_{i=1}^{s} \left|S(t_{i}) -S^o_{i} \right|^{2}, \\
        \text{MSE}_{I_{data}} &=\frac{1}{s} \sum_{i=1}^{s} \left|I(t_{i}) -I^o_{i} \right|^{2}, \\
        \text{MSE}_{R_{data}} &=\frac{1}{s} \sum_{i=1}^{s} \left|R(t_{i}) -R^o_{i} \right|^{2}. \\
    \end{aligned}
\end{equation}
%%%%%%%%%%%%%%%%%%%%%%%%%%%%%%%%%%%%%%%%%%%%%%%%%%%%%%%%%%%%%%%%%%%%%%%%%%%%%%%%%%%%%%%%%%
$S^o_{i},I^o_{i} $ and $R^o_{i}$ are the observed data at time points $t_{i}$, and $s$ is the number of observed data points. \par

The total loss of the present study consists of both, the data loss and the residual loss, according to which, both, the weights and biases, as well as the trainable dynamic parameters, are optimized,

%Then, in this study, the mean squared error (MSE) was coded as a loss function, which consisted of $\text{MSE}_{data}$ and $\text{MSE}_{SIR}$. We can determine the weights $\omega$ and biases $b$ as well as $\beta$ and $\gamma$ by solving the optimization problem
%%%%%%%%%%%%%%%%%%%%%%%%%%%%%%%%%%%%%%%%%%%%%%%%%%%%%%%%%%%%%%%%%%%%%%%%%%%%%%%%%%%%%%%%%%
\begin{equation}
        %\omega,b,\beta,\gamma = 
        \operatorname*{arg\,min}_{\omega, b,\beta,\gamma} (\text{MSE}_{data}+ \text{MSE}_{SIR}),
\end{equation}
%%%%%%%%%%%%%%%%%%%%%%%%%%%%%%%%%%%%%%%%%%%%%%%%%%%%%%%%%%%%%%%%%%%%%%%%%%%%%%%%%%%%%%%%%%
%also known as minimizing the loss function.\par

\subsection{Parameter Identifications}
\label{pia}
Algorithm~\ref{algorithm:1} shows how PINNs can be used to determine trainable parameters, including the NN- and SIR-model parameters. The input is the time point $t$, the output is the value of each compartment of the SIR model at a given $t$ value. Both, the weights $\omega$ and biases $b$, as well as the model parameters $\beta$ and $\gamma$, are initialized randomly ($\omega$ and $b$ use the PyTorch default initialization function, and $\beta$ and $\gamma$ were chosen randomly from (0,1). To ensure reproducibility, they were fixed during the practical experiments). 

% Algorithm~\ref{algorithm:1} shows how PINNs can be used to determine trainable parameters, including NN and SIR model parameters. The input is the time point $t$ and the output is the value of each compartment of the SIR model corresponding to $t$. The weights $\omega$ and biases $b$, as well as model parameters $\beta$ and $\gamma$ are initialized randomly ( $\omega$ and b use the PyTorch default initialization function, and $\beta$ and $\gamma$ use random numbers in the range 0 to 1).
%%%%%%%%%%%%%%%%%%%%%%%%%%%%%%%%%%%%%%%%%%%%%%%%%%%%%%%%%%%%%%%%%%%%%%%%%%%%%%%%%%%%%%%%%%
\begin{algorithm}[ht!]
  \caption{PINNs used to determine simultaneously the parameters of the neural network and the embedded SIR model.}
  \textbf{Data}: $t,S^o,I^o,R^o$ \\  Randomly initialize weights $\omega$, biases $b$, and dynamics parameters $\beta$, $\gamma$ \;
  %$\bm{Data}: t,S^o,I^o,R^o$ \\ $ Initialize \ weights \ w, \ biases \ b, \ and \ \beta, \ \gamma \ randomly $\;
    \For{\ epoch \ \textbf{in} \ epochs \ }
        {
            The values of each compartment of the SIR model can be obtained from the forward propagation of the neural network with the input as $t$
            %%%%%%%%%%%%%%%%%%%%%%%%%%%%%%%%%%%%%%%%%%%%%%%%%%%%%%%%%%%%%%%%%%%%%%%%%%%%%%
            \begin{equation}
                S,I,R = N\!N(t). \nonumber
            \end{equation}
            %%%%%%%%%%%%%%%%%%%%%%%%%%%%%%%%%%%%%%%%%%%%%%%%%%%%%%%%%%%%%%%%%%%%%%%%%%%%%%
            Evaluate the composed loss function, including the data loss (with $s$ to be the number of observations in each compartment, thus the number of time points collected):
            %%%%%%%%%%%%%%%%%%%%%%%%%%%%%%%%%%%%%%%%%%%%%%%%%%%%%%%%%%%%%%%%%%%%%%%%%%%%%%
            \begin{equation}
                \text{MSE}_{SIR} =\frac{1}{s} \sum_{i=1}^{s}(\left|S_{i}-S^o_{i}\right|^{2}+\left|I_{i}-I^o_{i}\right|^{2}+\left|R_{i}-R^o_{i}\right|^{2}), \nonumber
            \end{equation}
            %%%%%%%%%%%%%%%%%%%%%%%%%%%%%%%%%%%%%%%%%%%%%%%%%%%%%%%%%%%%%%%%%%%%%%%%%%%%%%
            denoting the mismatch of the output of the neural network and observation data. Here the residual loss:
            %%%%%%%%%%%%%%%%%%%%%%%%%%%%%%%%%%%%%%%%%%%%%%%%%%%%%%%%%%%%%%%%%%%%%%%%%%%%%%
            \begin{small}
            \begin{equation}
                \text{MSE}_{Residuals}=\frac{1}{q} \sum_{i=1}^{q}\left(\left|\frac{d {S}_{i}}{d t_{i}}+\frac{\beta {S}_{i} {I}_{i}} {N} \right|^{2} 
                + \left|\frac{d {I}_{i}}{d t_{i}}-\frac{\beta {S}_{i} {I}_{i}} {N} + \gamma I_{i} \right|^{2}
                + \left|\frac{d {R}_{i}}{d t_{i}}- \gamma I_{i} \right|^{2} \right
                ) \nonumber
            \end{equation}
            \end{small}
            %%%%%%%%%%%%%%%%%%%%%%%%%%%%%%%%%%%%%%%%%%%%%%%%%%%%%%%%%%%%%%%%%%%%%%%%%%%%%%
            stands for the sum of the residual errors for each compartment of the SIR model. Here, the residuals and data loss are calculated using the same time step $\Delta t=1$. Thus, the total loss function can be obtained:
            %%%%%%%%%%%%%%%%%%%%%%%%%%%%%%%%%%%%%%%%%%%%%%%%%%%%%%%%%%%%%%%%%%%%%%%%%%%%%%
            \begin{equation}
                Loss = \text{MSE}_{SIR}+\text{MSE}_{Residuals} \nonumber
            \end{equation}
            %%%%%%%%%%%%%%%%%%%%%%%%%%%%%%%%%%%%%%%%%%%%%%%%%%%%%%%%%%%%%%%%%%%%%%%%%%%%%%
            The Adam Optimizer\upcite{Kingma2014adam} toolkit in Pytorch is utilized to update the weights $\omega$ and biases $b$, as well as $\beta$ and $\gamma$ by minimizing the loss function.
        }
    \label{algorithm:1}
    \end{algorithm}
%%%%%%%%%%%%%%%%%%%%%%%%%%%%%%%%%%%%%%%%%%%%%%%%%%%%%%%%%%%%%%%%%%%%%%%%%%%%%%%%%%%%%%%%%%
% \newpage

\section{Experiments and Results}
\label{sec:res}
 This section presents first the setup for generating the synthetic data from the SAIRD model, and then the pre-processing for the data reported for Germany. The details of the here devised physics-informed neural networks for the above two situations follow. Finally, the performance of proposed the PINNs are demonstrated for the testing data sets.\par

 % In this section, we first present the setup of generating synthetic data from the SAIRD model, and the pre-processing for the reported data of Germany. Next, the details about the devised physics-informed neural networks for the above two situations will be given. Finally, the performance of proposed PINNs will be demonstrated on the testing data sets.\par

% In this section, we first present the setup of the neural network used for the experiments and mathematical models to generate the synthetic data. Next, we describe the pre-processing methods for the data. Finally, we present and discuss the proposed method's results on synthetic and realistic data from Germany.\par
\subsection{Data Preparation}
\subsubsection{Synthetic data generation and processing}
\begin{figure}[ht!]
	\centering
	\includegraphics[width=14cm]{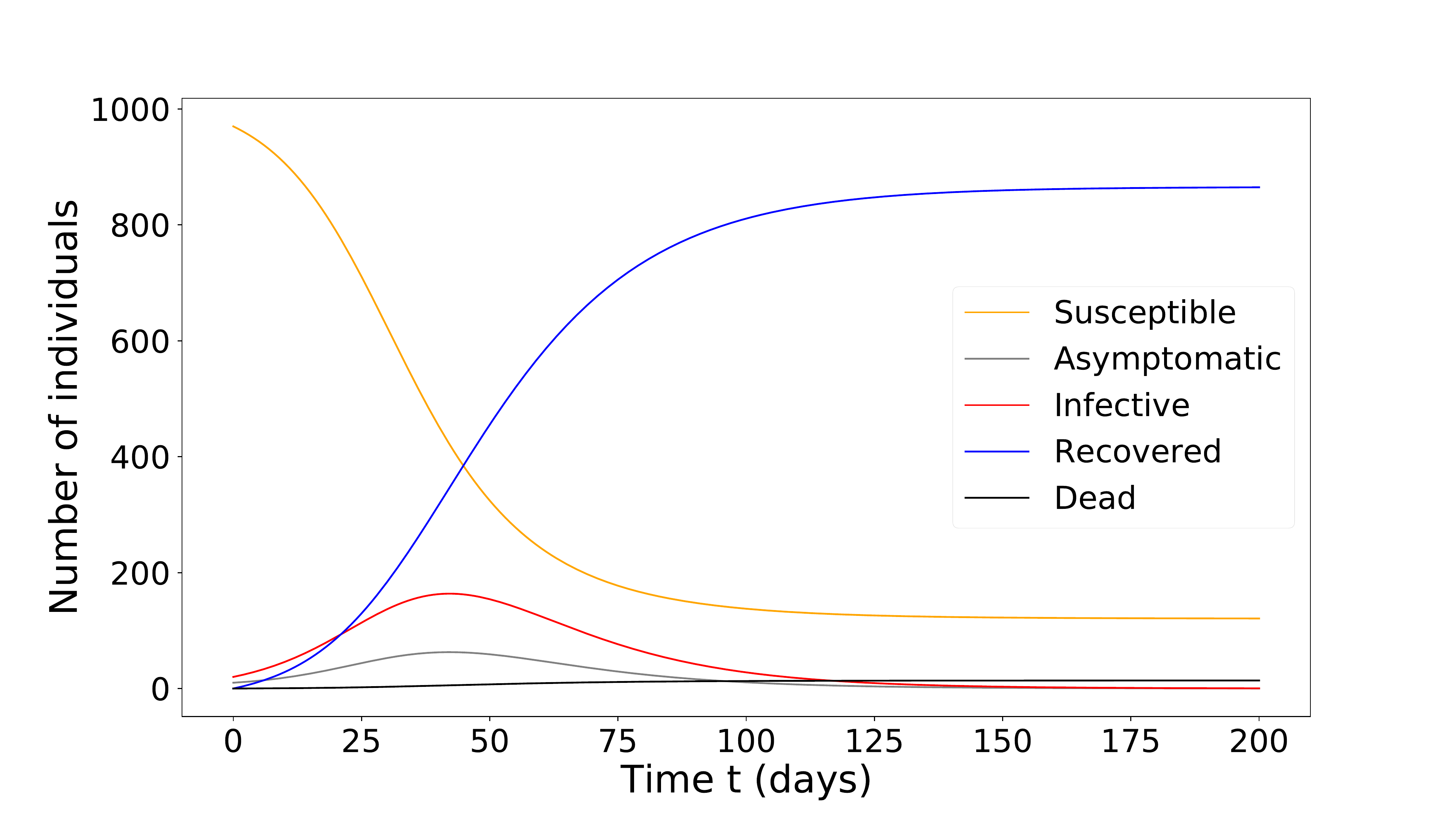}	
	\caption{A Mathematical model generated time evolution of an example SAIRD model. The solid yellow line represents the number of susceptible people in the population, the solid grey line represents the number of asymptomatic infected people, the solid blue line represents the number of recovered people, and the solid red line stand for the number of active infected person. The solid black line is the death population, and The population is assumed to be constant (N=1000).}
	\label{fig:saird_data}
\end{figure}

To test and validate the PINN frameworks, whose physical laws are derived from the simple SIR model, a complex mathematical model, the SAIRD model \eqref{equ:saird} is used, to generate the mock data. The advantage of using this type of data is that here the dynamics is clean, little noisy, verify that the approach does work. The data shown in Figure~\ref{fig:saird_data} starting the model evolution with the given model parameters and the initial conditions as shown in Table~\ref{tab:parameters}. \par

% To test and validate the PINN framework, whose physical laws are derived from the simple SIR model, we used a more complex mathematical model, the SAIRD model \eqref{equ:saird}, to generate mock data. The advantage of this type of data is that its dynamics is clean and less noisy for the purpose of verification of the approach. In obtaining the data shown in Figure~\ref{fig:saird_data}, we start the model with the given model parameters and initial conditions shown in Table~\ref{tab:parameters}. \par
%%%%%%%%%%%%%%%%%%%%%%%%%%%%%%%%%%%%%%%%%%%%%%%%%%%%%%%%%%%%%%%%%%%%%%%%%%%%%%%%%%%%%%%%%%
\begin{table}[ht!]
\scriptsize
\centering
\caption{The value of parameters and initial conditions for the SAIRD model.}
\begin{tabular}{ccc} 
    \toprule
        Parameter     & Description                                                        & Value \\ \midrule
        $\rho_{1}$    & Probability that one becoming symptomatically by exposure to an asymptomatic carrier                         & 0.80  \\
        $\rho_{2}$    & Probability that one becoming asymptomatically by exposure to a symptomatically carrier                       & 0.29  \\
        $\alpha$      & Infection rate of individual exposed to an asymptomatic carrier    & 0.1   \\
        $\beta$       & Infection rate of individual exposed to symptomatic carrier        & 0.17  \\
        $\gamma$      & Recovery rate                                                      & 1/16  \\
        $\theta$      & Death rate                                                         & 0.001 \\
        $N$           & Population                                                         & 1000  \\
        $A_{0}$       & Initial number of Asymptomatic infectious individuals              & 10    \\
        $I_{0}$       & Initial number of Infectious individuals                           & 20    \\
        $R_{0}$       & Initial number of Recovered individuals                            & 0     \\
        $D_{0}$       & Initial number of Dead individuals                                & 0     \\
        $S_{0}$       & Initial number of Susceptible individuals                          & 970   \\
    \bottomrule
    \end{tabular}
    \label{tab:parameters}
\end{table}
%%%%%%%%%%%%%%%%%%%%%%%%%%%%%%%%%%%%%%%%%%%%%%%%%%%%%%%%%%%%%%%%%%%%%%%%%%%%%%%%%%%%%%%%%%%
Figure~\ref{fig:saird_data} shows the time evolution of the classical SIR model. Here the number of susceptibles in the population decreases as the number of actively infected people increases. The population size for recovered and death (removal) then increases until it hits a maximum or stabilizes. \par
% %%%%%%%%%%%%%%%%%%%%%%%%%%%%%%%%%%%%%%%%%%%%%%%%%%%%%%%%%%%%%%%%%%%%%%%%%%%%%%%%%%%%%%%%%%%
% \begin{figure}[ht!]
% 	\centering
% 	\includegraphics[width=12cm]{figures/synthetic_data/saird_model_generateion_data.pdf}	
% 	\caption{A Mathematical model generated time evolution of an example SAIRD model. The solid yellow line represents the number of susceptible people in the population, the solid grey line represents the number of asymptomatic infected people, the solid blue line represents the number of recovered people, and the solid red line stand for the number of active infected person. The solid black line is the death population, and The population is assumed to be constant (N=1000).}
% 	\label{fig:saird_data}
% \end{figure}
% %%%%%%%%%%%%%%%%%%%%%%%%%%%%%%%%%%%%%%%%%%%%%%%%%%%%%%%%%%%%%%%%%%%%%%%%%%%%%%%%%%%%%%%%%%%

%%%%%%%%%%%%%%%%%%%%%%%%%%%%%%%%%%%%%%%%%%%%%%%%%%%%%%%%%%%%%%%%%%%%%%%%%%%%%%%%%%%%%%%%%%%
%%%%%%%%%%%%%%%%%%%%%%%%%%%%%%%%%%%%%%%%%%%%%%%%%%%%%%%%%%%%%%%%%%%%%%%%%%%%%%%%%%%%%%%%%%%

The epidemiological characteristics of the SAIRD model, both its S (Susceptible) and A (Asymptomatic) compartments are derived from the S (Susceptible) compartment of the SIR model. The SAIRD's R (Recovered) and D (Dead) compartments are obtained from the R (Removed) compartment in the SIR model. 

% In part B of the Appendix, Figure~\ref{fig:saird_data} shows the behaviour of the classical SIR model over time, with the number of susceptible populations decreasing as the scale of active disease increases. The population scale for recovery and death (removal) then increases until it hits a maximum or stabilizes. \par

\begin{figure}[ht!]
	\centering
	\includegraphics[width=14cm]{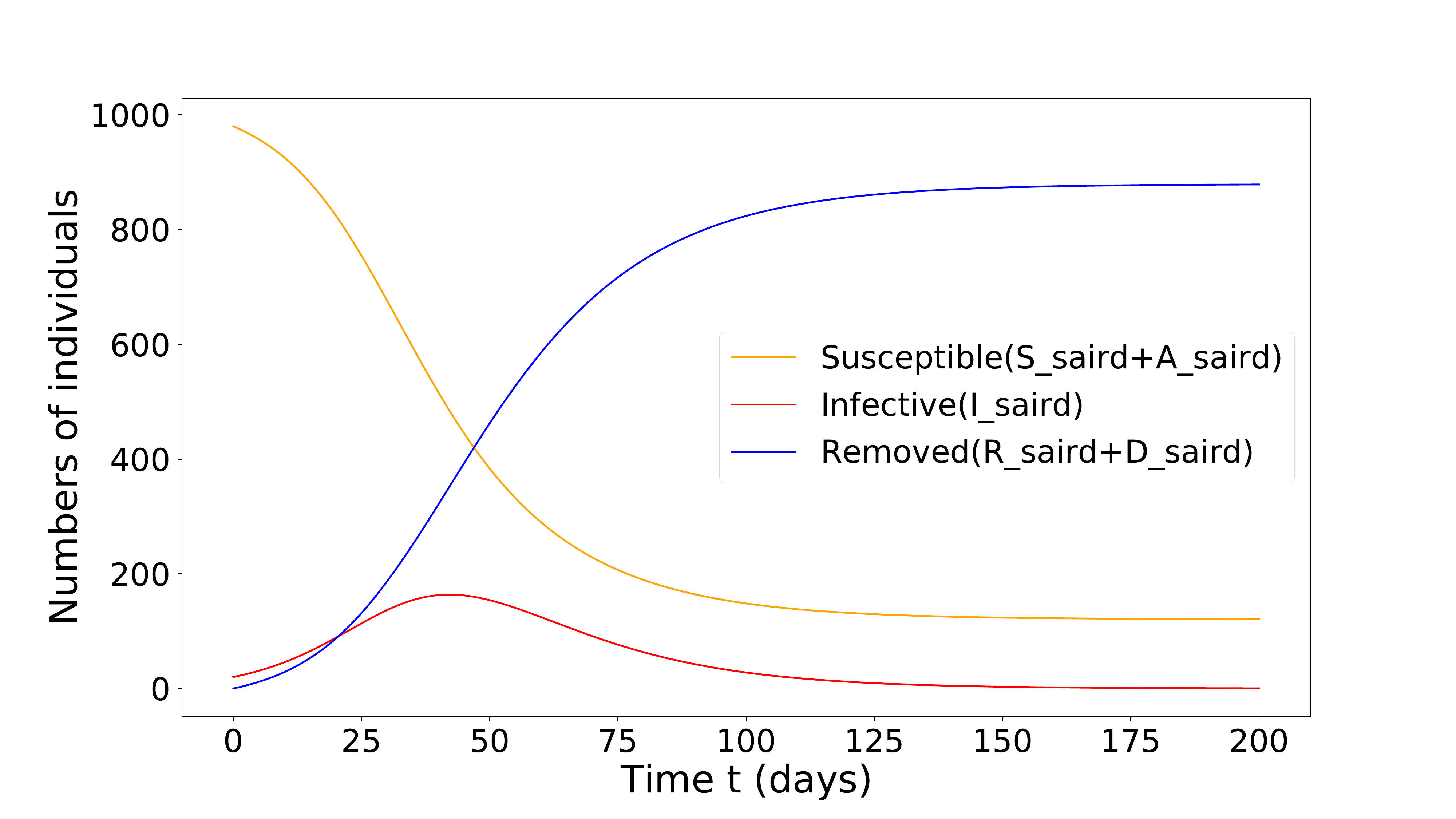}
	\caption{SIR mock data derived from SAIRD model generation data.}
	\label{fig:sir_data}
\end{figure}

% In addition, due to the epidemiological characteristics of the SAIRD model, its S (Susceptible) and A (Asymptomatic) compartments are derived from the S (Susceptible) compartment in the SIR model. On the other hand, its R (Recovered) and D (Death) compartments are derived from the R (Removed) compartment in the SIR model.
%%%%%%%%%%%%%%%%%%%%%%%%%%%%%%%%%%%%%%%%%%%%%%%%%%%%%%%%%%%%%%%%%%%%%%%%%%%%%%%%%%%%%%%%%%%
% \begin{figure}[ht!]
% 	\centering
% 	\includegraphics[width=12cm]{figures/synthetic_data/saird_model_generateion_data.pdf}	
% 	\caption{A Mathematical model generated graph simulation of an example SAIRD model. The solid yellow line represents a susceptible population, a solid grey line represents the asymptomatic, a solid blue line represents the recoveries, and the solid red line is the active infected population. The solid black line is the deceased population, and The population is assumed to be constant (N=1000).}
% 	\label{fig:saird_data}
% \end{figure}
% %%%%%%%%%%%%%%%%%%%%%%%%%%%%%%%%%%%%%%%%%%%%%%%%%%%%%%%%%%%%%%%%%%%%%%%%%%%%%%%%%%%%%%%%%%%
Hence, the values of the S and R compartments of the SIR model are obtained by adding the values of A to S in the SAIRD model, and the values of R to D, respectively. The compartment I have been defined identically in both models. So, the values of the three compartments in the SIR model, shown in Figure~\ref{fig:sir_data} are used as simulation data for the straightforward tests and validations to follow here.
% %%%%%%%%%%%%%%%%%%%%%%%%%%%%%%%%%%%%%%%%%%%%%%%%%%%%%%%%%%%%%%%%%%%%%%%%%%%%%%%%%%%%%%%%%%%
% \begin{figure}[ht!]
% 	\centering
% 	\includegraphics[width=12cm]{figures/synthetic_data/sir_from_saird_model_generateion_data.pdf}
% 	\caption{SIR mock data derived from SAIRD model generation data.}
% 	\label{fig:sir_data}
% \end{figure}
% %%%%%%%%%%%%%%%%%%%%%%%%%%%%%%%%%%%%%%%%%%%%%%%%%%%%%%%%%%%%%%%%%%%%%%%%%%%%%%%%%%%%%%%%%%

%%%%%%%%%%%%%%%%%%%%%%%%%%%%%%%%%%%%%%%%%%%%%%%%%%%%%%%%%%%%%%%%%%%%%%%%%%%%%%%%%%%%%%%%%%%
%%%%%%%%%%%%%%%%%%%%%%%%%%%%%%%%%%%%%%%%%%%%%%%%%%%%%%%%%%%%%%%%%%%%%%%%%%%%%%%%%%%%%%%%%%

\begin{figure}[ht!]
	\centering
	\includegraphics[width=14cm]{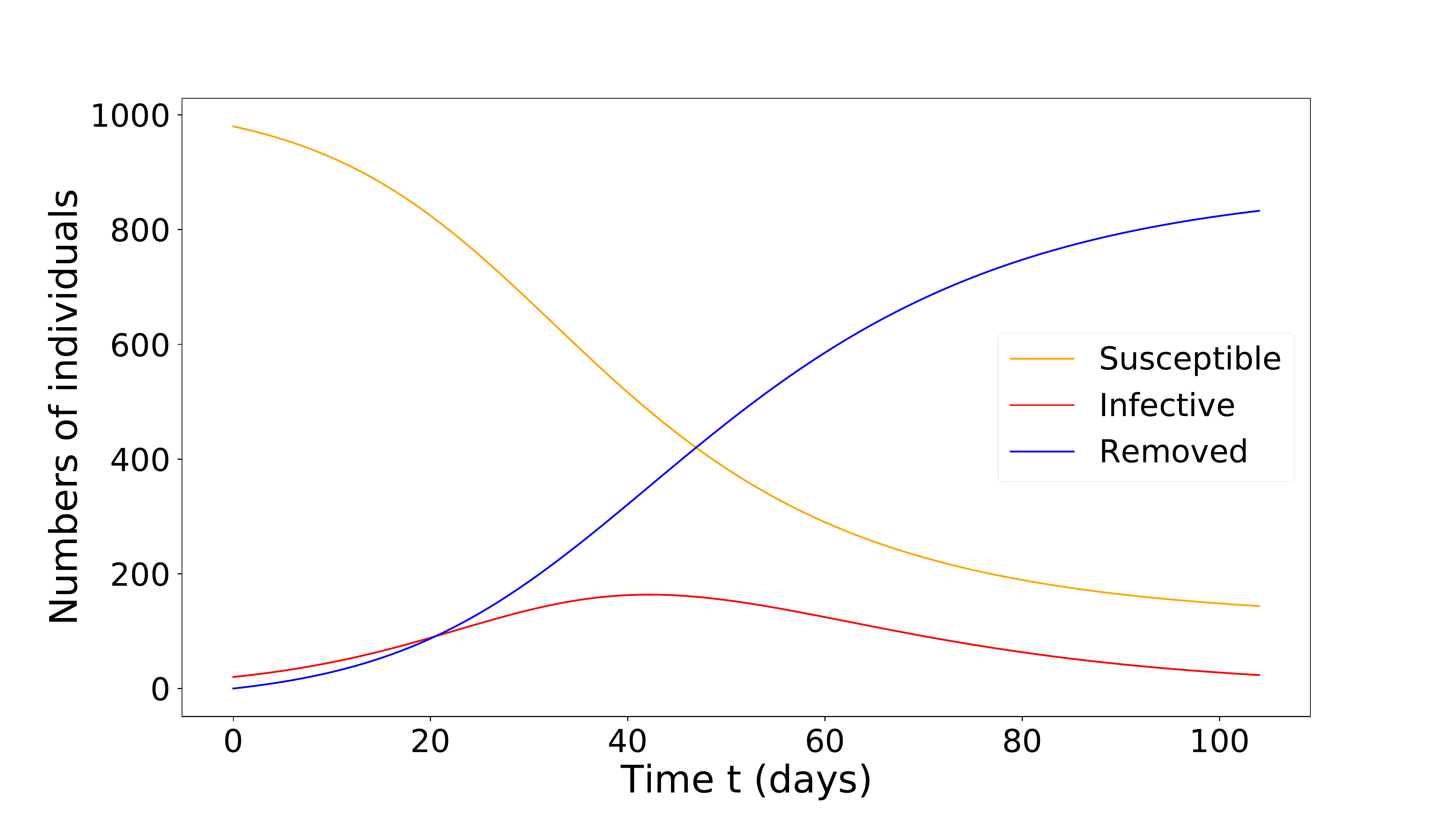}
	\caption{Selected SIR mock data derived from the SAIRD model generation data are used for the synthetic data simulations.}
	\label{fig:sir_selected_data}
\end{figure}

% Therefore, we can obtain the values of the S and R compartments in the SIR model by adding the value of A to S in the SAIRD model and the value of R to D, respectively. Moreover, the I compartment has been defined identically in both models; we can obtain the values of the three compartments in the SIR model, as shown in Figure~\ref{fig:sir_data} in Appendix B, as simulation data for the following tests and validations.
% %%%%%%%%%%%%%%%%%%%%%%%%%%%%%%%%%%%%%%%%%%%%%%%%%%%%%%%%%%%%%%%%%%%%%%%%%%%%%%%%%%%%%%%%%%%
% \begin{figure}[ht!]
% 	\centering
% 	\includegraphics[width=12cm]{figures/synthetic_data/sir_from_saird_model_generateion_data.pdf}
% 	\caption{SIR mock data derived from SAIRD model generation data.}
% 	\label{fig:sir_data}
% \end{figure}
% %%%%%%%%%%%%%%%%%%%%%%%%%%%%%%%%%%%%%%%%%%%%%%%%%%%%%%%%%%%%%%%%%%%%%%%%%%%%%%%%%%%%%%%%%%
All values of all compartments stabilize after approximately 100 days, the first 100 days, where compartment I exhibits a broad, 30 day wide peak, are chosen as the synthetic dataset for the calculations and testing shown in Figure~\ref{fig:sir_selected_data}.
% In addition, as the values of all compartments stabilize after 100 days, the first 100 days in which the I compartment covers a peak, were chosen as the synthetic dataset for easy calculation and testing, as shown in Figure~\ref{fig:sir_selected_data}.
% %%%%%%%%%%%%%%%%%%%%%%%%%%%%%%%%%%%%%%%%%%%%%%%%%%%%%%%%%%%%%%%%%%%%%%%%%%%%%%%%%%%%%%%%%%
% \begin{figure}[ht!]
% 	\centering
% 	\includegraphics[width=12cm]{figures/synthetic_data/sir_selected_saird_model_generateion_data.pdf}
% 	\caption{Selected SIR mock data derived from the SAIRD model generation data are used for the synthetic data simulations.}
% 	\label{fig:sir_selected_data}
% \end{figure}
% %%%%%%%%%%%%%%%%%%%%%%%%%%%%%%%%%%%%%%%%%%%%%%%%%%%%%%%%%%%%%%%%%%%%%%%%%%%%%%%%%%%%%%%%%%

%%%%%%%%%%%%%%%%%%%%%%%%%%%%%%%%%%%%%%%%%%%%%%%%%%%%%%%%%%%%%%%%%%%%%%%%%%%%%%%%%%%%%%%%%%
%%%%%%%%%%%%%%%%%%%%%%%%%%%%%%%%%%%%%%%%%%%%%%%%%%%%%%%%%%%%%%%%%%%%%%%%%%%%%%%%%%%%%%%%%%

\subsubsection{Pre-processing of reported data}
The COVID-19 epidemic data of Germany are used for our model analysis. they are based on the cumulative number of infected individuals, the cumulative recovered individuals, and the cumulative number of deaths, as reported by Robert Koch-Institut (RKI) COVID-19 Data for the period from March 1 to July 1, 2021. The SIR dynamics integrated into our method. Hence, the first step of the preprocessing of the reported data is to extract the values for the different compartment from the records. The data for Germany track the cumulative infectious, $I_{c}$, recovered, $R_{c}$, and deceased cases, $D_{c}$. In accord with the definition of the compartment $R$, in the SIR model, $R$ contains both the recovered as well as the virus-induced dead individuals, $R=R_{recovered}+D_{death}$. As the compartment $R$ does not have any outflow, $R=Re_{c}$. Here, $Re_{c}$ is the cumulative value of the removed. Given the reported data's values, the theoretical values for removed, infectious, susceptible cases are obtained by $R=R_{recovered}+D_{death}$, $I = I_{c} -R$, and $S = N {-} I {-} R$. Here, $N$ is the total population. All values of all compartments are normalized by dividing by the total population. Thus, N is set to 1.\par

% We consider reported data of the COVID-19 epidemic in Germany for our analysis, based on the cumulative number of infected individuals, cumulative recovered individuals, and the cumulative number of deaths reported by JHU CSSE COVID-19 Data\upcite{dong2020interactive} for the period from 1 March to 1 July 2021. As the SIR dynamics is integrated into our proposed method, the first step of preprocessing the reported data is to extract values for different compartment from the record. The German data tracks the cumulative infectious $I_{c}$, recovered $R_{c}$, and deceased cases $D_{c}$. According to the definition of compartment $R$ in the SIR model, $R$ contains recovered and virus-induced dead individuals, so $R=R_{recovered}+D_{death}$, and because compartment $R$ does not have any outflow, $R=Re_{c}$ where $Re_{c}$ is the cumulative removed value. Given the reported values, we can obtain the theoretical values for removed, infectious, susceptible cases by $R=R_{recovered}+D_{death}$, $I = I_{c} -R$, and $S = N {-} I {-} R$, where $N$ is the total population. As normalization, the values for all compartments are divided by the total population so N is set to 1.\par

\begin{figure}[ht!]
    \centering
	\includegraphics[width=0.95\textwidth]{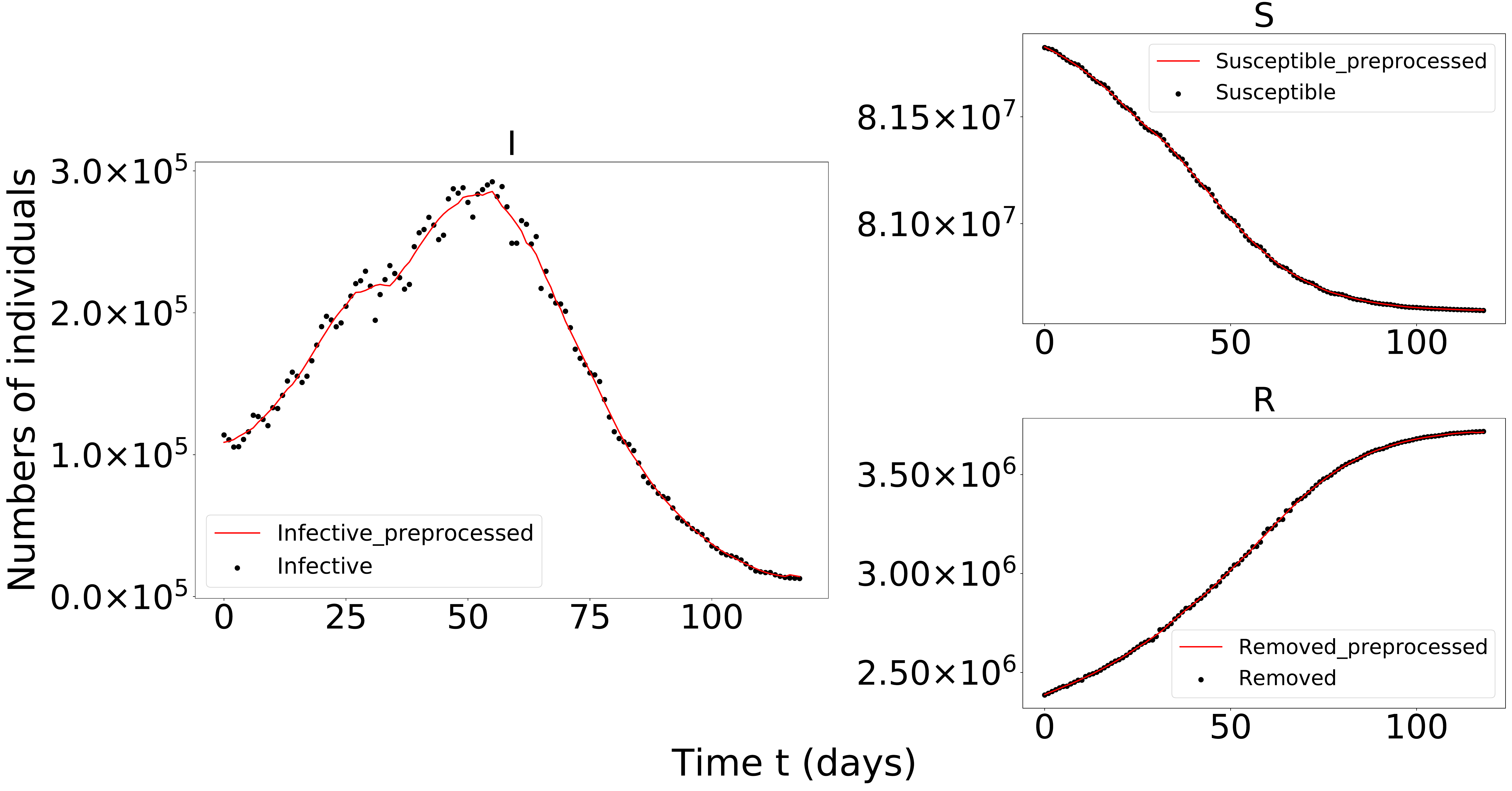}
 
    \caption{The figure compares the results before and after pre-processing of the reported data. I, S and R correspond to the trends in the number of infectious, susceptible, and removed individuals, respectively, where the black scatter shows the raw, unprocessed data, and the solid red line is a curve of the data processed through a sliding window. The graphs show that the pre-processed data curves are smoother, and the fluctuations are reduced effectively.}
    
    \label{fig:reported_data.processed}

\end{figure}

Reliability and cleanliness of the reported data matter a lot for the predictability of compartmental models, in particular for the sense of parameter estimates. Reported data are, however, quite noisy, both due to misreporting, late reporting and other reasons. The reported data from Germany vary greatly substantially between weekdays and weekends. There is due to the significant reduction in detection at weekends. Therefore, the ability to make valuable estimates from the available reported data may be limited. Thus, before the data was analysed, we pre-process the dataset by applying a 7-day moving average window to smooth out weekday-weekend zigzags in the outbreak reports. As shown in the Figure~\ref{fig:reported_data.processed}, the reported data from the different compartments are significantly smoother and less noisy than the raw data after the sliding window pre-processing. \par

\subsection{Setup of SIR-Infomed Neural Networks}
\label{sec: snn}

The neural network structure we use in our experiments takes a single value, the time $t$, as input. There're weights $W_{i,j}$ being associated for each hidden layer, where $i$ is the position of the start node and $j$ is the position of the end node. The non-linear activation function $tanh$ is applied at each node in the hidden layer, %The product of the weights $W_{i,j}$ and time $t$ is applied at each node to the activation function $tanh$
%%%%%%%%%%%%%%%%%%%%%%%%%%%%%%%%%%%%%%%%%%%%%%%%%%%%%%%%%%%%%%%%%%%%%%%%%%%%%%%%%%%%%%%%%%
\begin{equation}
        \tanh{(x)}=\frac{e^{x}-e^{-x}}{e^{x}+e^{-x}}.
\end{equation}
%%%%%%%%%%%%%%%%%%%%%%%%%%%%%%%%%%%%%%%%%%%%%%%%%%%%%%%%%%%%%%%%%%%%%%%%%%%%%%%%%%%%%%%%%%
For the output nodes of the network, $sigmoid$ activation function is applied, 
%%%%%%%%%%%%%%%%%%%%%%%%%%%%%%%%%%%%%%%%%%%%%%%%%%%%%%%%%%%%%%%%%%%%%%%%%%%%%%%%%%%%%%%%%%
\begin{equation}
        \sigma(x)=\frac{1}{1+e^{-x}},
\end{equation}
%%%%%%%%%%%%%%%%%%%%%%%%%%%%%%%%%%%%%%%%%%%%%%%%%%%%%%%%%%%%%%%%%%%%%%%%%%%%%%%%%%%%%%%%%%
to account for the normalization applied to $S(t)$, $I(t)$ and $R(t)$ respectively. The neural network contains 4 hidden layers and 64 neurons in each layer. The Adam algorithm of the PyTorch package was selected as the optimizer. The used learning rate is 0.0001 and the training epochs is 200k. To prevent overfitting and improve the generalization ability of the model in confronting new data, we applied early stopping, to which the epoch limit is set to be 300.\par

\subsection{Performance on synthetic data}
\label{sec: md}

%We chose 15, 25 and 37 days before the peak of compartment I as the training set to test and validate the proposed method. 
% First, the PINNs learned parameters are used to produce the numerical solution of SIR model as Eq.\eqref{equ:sir} shows, which is obtained by using the LSODA algorithm\lw{need at least Ref.}, which is used to solve the numerical solution problem for rigid or non-rigid systems of first-order ordinary differential equations. Then, the solution of the ODEs system is compared with the results by PINNs.\par

%%%%%%%%%%%%%%%%%%%%%%%%%%%%%%%%%%%%%%%%%%%%%%%%%%%%%%%%%%%%%%%%%%%%%%%%%%%%%%%%%%%%%%%%%
\begin{figure}[ht!]
    \centering
\includegraphics[width=0.95\textwidth]{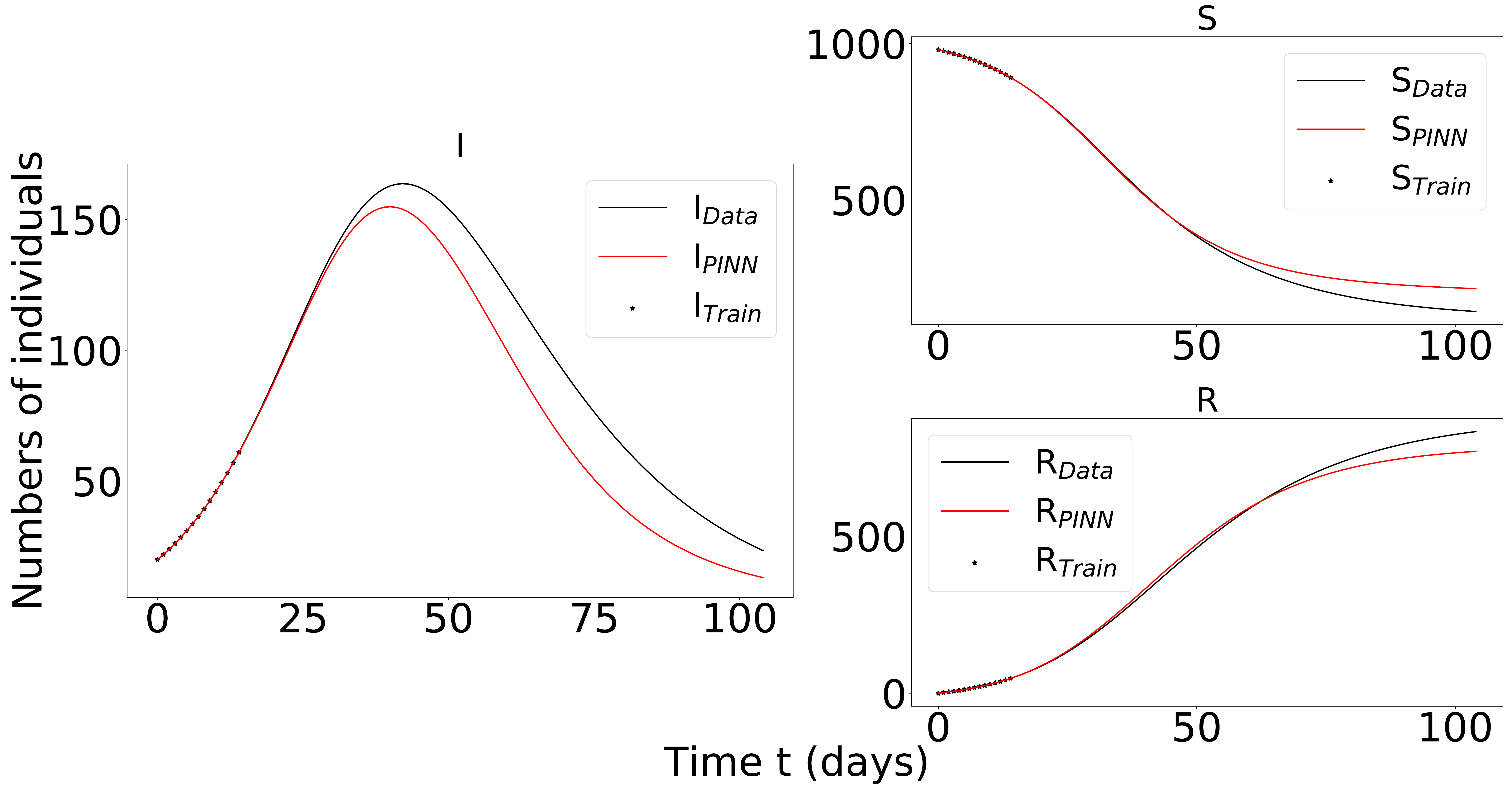}

    \caption{The graph shows results identified and predicted using the first 15 days as the training set. I, S and R diagrams of compartments I, S and R, respectively. The black star indicates the training set, the solid black line is the synthetic data, the solid red line is the result from PINNs.}
    \label{fig.mockdata.15}

\end{figure}
%%%%%%%%%%%%%%%%%%%%%%%%%%%%%%%%%%%%%%%%%%%%%%%%%%%%%%%%%%%%%%%%%%%%%%%%%%%%%%%%%%%%%%%%%%
\par
%%%%%%%%%%%%%%%%%%%%%%%%%%%%%%%%%%%%%%%%%%%%%%%%%%%%%%%%%%%%%%%%%%%%%%%%%%%%%%%%%%%%%%%%%%
\begin{figure}[ht!]
    \centering
    \includegraphics[width=0.95\textwidth]{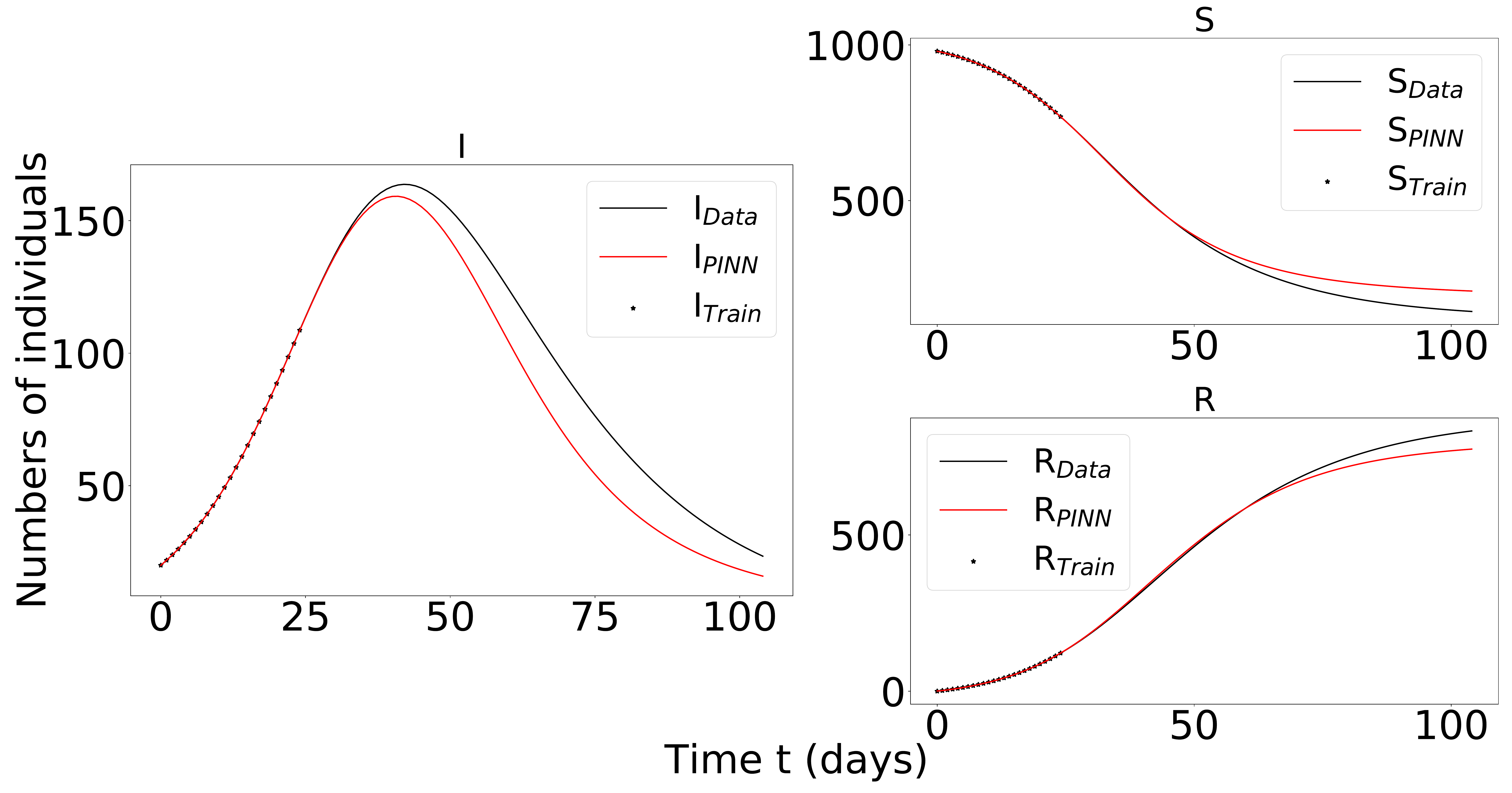}

    \caption{The graph shows results identified and predicted using the first 25 days as the training set. I, S and R diagrams of compartments I, S and R, respectively. The black star indicates the training set, the solid black line is the synthetic data, the solid red line is the result from PINNs.}
    \label{fig.mockdata.25}

\end{figure}
%%%%%%%%%%%%%%%%%%%%%%%%%%%%%%%%%%%%%%%%%%%%%%%%%%%%%%%%%%%%%%%%%%%%%%%%%%%%%%%%%%%%%%%%%%
\par
%%%%%%%%%%%%%%%%%%%%%%%%%%%%%%%%%%%%%%%%%%%%%%%%%%%%%%%%%%%%%%%%%%%%%%%%%%%%%%%%%%%%%%%%%%

\begin{figure}[ht!]
    \centering
    \includegraphics[width=0.95\textwidth]{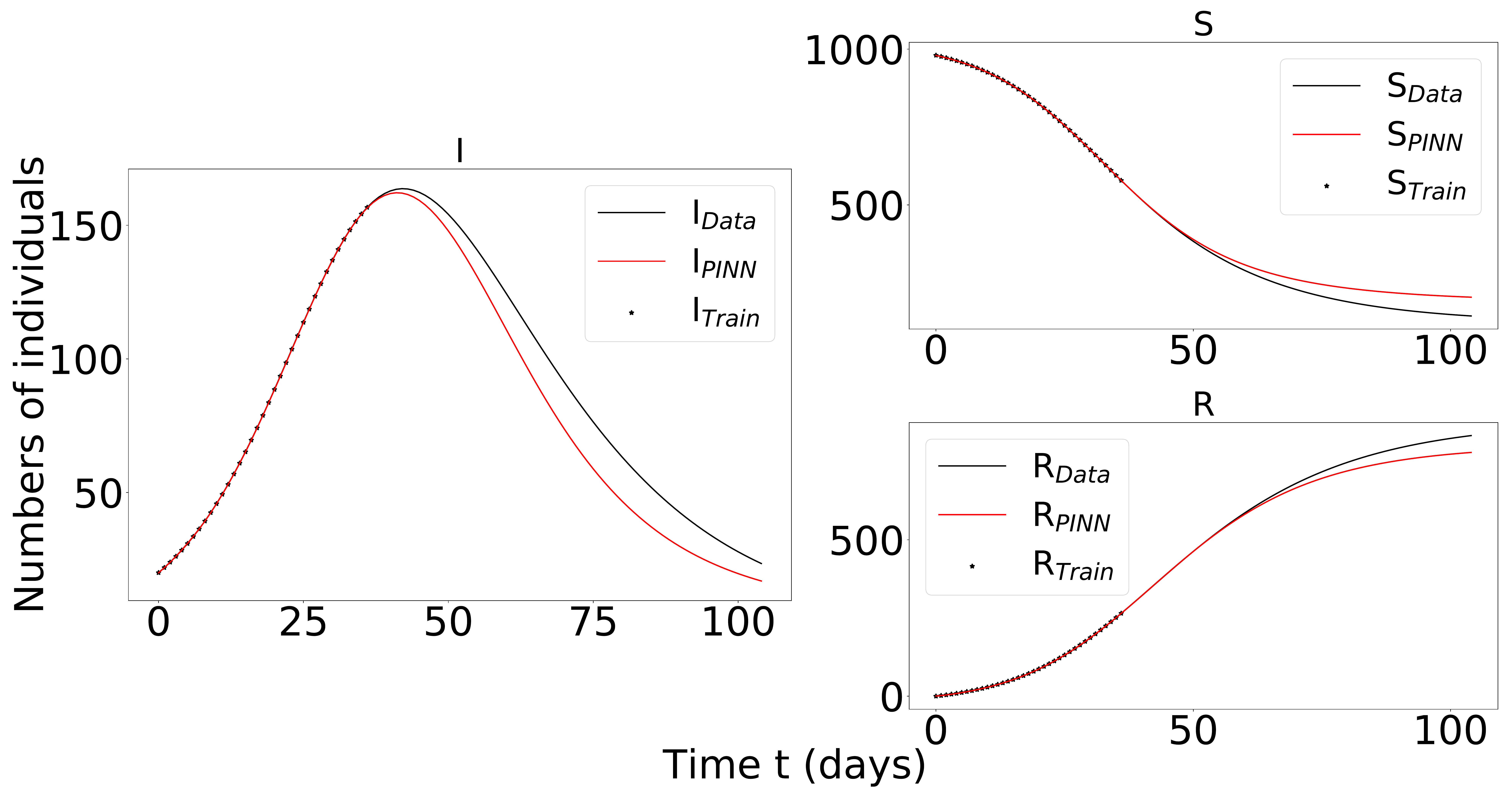}

    \caption{The graph shows results identified and predicted using the first 37 days as the training set. I, S and R diagrams of compartments I, S and R, respectively. The black star indicates the training set, the solid black line is the synthetic data, the solid red line is the result from PINNs.}
    \label{fig.mockdata.37}

\end{figure}

% \begin{figure}[hbtp!]

%     \begin{minipage}{.65\linewidth}
%         \centering
%         \subfloat[][I]{
%             \label{mockdata.37:I}
%             \includegraphics[width=.95\linewidth]{figures/synthetic_data/37_for_I_mockdata.pdf}}
%     \end{minipage}
%     \medskip
%     \begin{minipage}{.35\linewidth}
%         \centering
%         \subfloat[][S]{
%             \label{mockdata.37:S}
%             \includegraphics[width=.95\linewidth]{figures/synthetic_data/37_for_S_mockdata.pdf}}
    
%         \subfloat[][R]{
%             \label{mockdata.37:R}
%             \includegraphics[width=.95\linewidth]{figures/synthetic_data/37_for_R_mockdata.pdf}}
%     \end{minipage} 

%     \caption{The graph shows results identified and predicted using the first 37 days as the training set. (a), (b) and (c) diagrams of compartments I, S and R, respectively. The black star indicates the training set, the solid black line is the synthetic data, the solid red line is the result from PINNs.}
%     \label{fig.mockdata.37}

% \end{figure}
%%%%%%%%%%%%%%%%%%%%%%%%%%%%%%%%%%%%%%%%%%%%%%%%%%%%%%%%%%%%%%%%%%%%%%%%%%%%%%%%%%%%%%%%%%
% todo mock data about SIR-ODEs
We investigated using respectively 15, 25 and 37 days of data as the training set and, to ensure reproducibility, the same initial values of $\beta = 0.15$ and $\gamma = 0.15$ to test the identifiability and predictability of our proposed approach. In Figures~\ref{fig.mockdata.15}, \ref{fig.mockdata.25} and \ref{fig.mockdata.37}, we present identification and prediction results of our proposed method for using different sizes of training sets on mock data. Clearly we see that for a complex system including compartments S and R, the PINNs integrated with simple SIR dynamics can fit and predict well on the training and test sets. The numerical solutions for the ODE with the PINN learned parameters also perform well which basically would overlap the PINN results. However, for data with peaks like compartment I, the SIR dynamics informed neural network model gives slightly better results than the numerical solution of the ODE, especially at the peaks. This can also be seen in the loss value shown in the Table~\ref{tab:error} of the results. \par
%%%%%%%%%%%%%%%%%%%%%%%%%%%%%%%%%%%%%%%%%%%%%%%%%%%%%%%%%%%%%%%%%%%%%%%%%%%%%%%%%%%%%%%%%%
\begin{table}[hbtp!]
\centering
\caption{Error Metrics for PINNs and SIR-ODEs on different training set size.}
\label{tab:error}
\begin{tabular}{@{}cccc@{}}
\toprule
    \multicolumn{2}{c}{}       &MSE                &MAE       \\ \midrule
    \multirow{2}*{15} &PINNs &2202.6 &53.889  \\
                      &SIR-ODEs  &2946.3  &63.172 \\
    \multirow{2}*{25} &PINNs &1790.6 &46.215   \\ 
                      &SIR-ODEs  &2504.0 &57.002 \\  
    \multirow{2}*{37} &PINNs &1461.8 &39.919 \\
                      &SIR-ODEs  &2025.6 &51.145  \\ \midrule 
\end{tabular}
\end{table}   
%%%%%%%%%%%%%%%%%%%%%%%%%%%%%%%%%%%%%%%%%%%%%%%%%%%%%%%%%%%%%%%%%%%%%%%%%%%%%%%%%%%%%%%%%%
\par
%%%%%%%%%%%%%%%%%%%%%%%%%%%%%%%%%%%%%%%%%%%%%%%%%%%%%%%%%%%%%%%%%%%%%%%%%%%%%%%%%%%%%%%%%%
\begin{table}[hbtp!]
\centering
\caption{Parameter learned for PINNs on different training set size.}
\label{tab:parameters_earned}
\begin{tabular}{@{}ccccc@{}}
\toprule
\multirow{2}*{} &  & \multicolumn{3}{c}{Parameters learned} \\ \cmidrule(l){2-5} 
               & initial & 15          & 25          & 37         \\ \midrule
$\beta$        & 0.15    & 0.18266      & 0.17948      & 0.17689  \\
$\gamma$       & 0.15    & 0.09486      & 0.09208      & 0.08948  \\ \midrule  
\end{tabular}
\end{table}
%%%%%%%%%%%%%%%%%%%%%%%%%%%%%%%%%%%%%%%%%%%%%%%%%%%%%%%%%%%%%%%%%%%%%%%%%%%%%%%%%%%%%%%%%%

\subsection{Performance on Reported Germany Data}
\label{sec: rd}
\begin{figure}[ht!]
	\centering
    \includegraphics[width=0.95\textwidth]{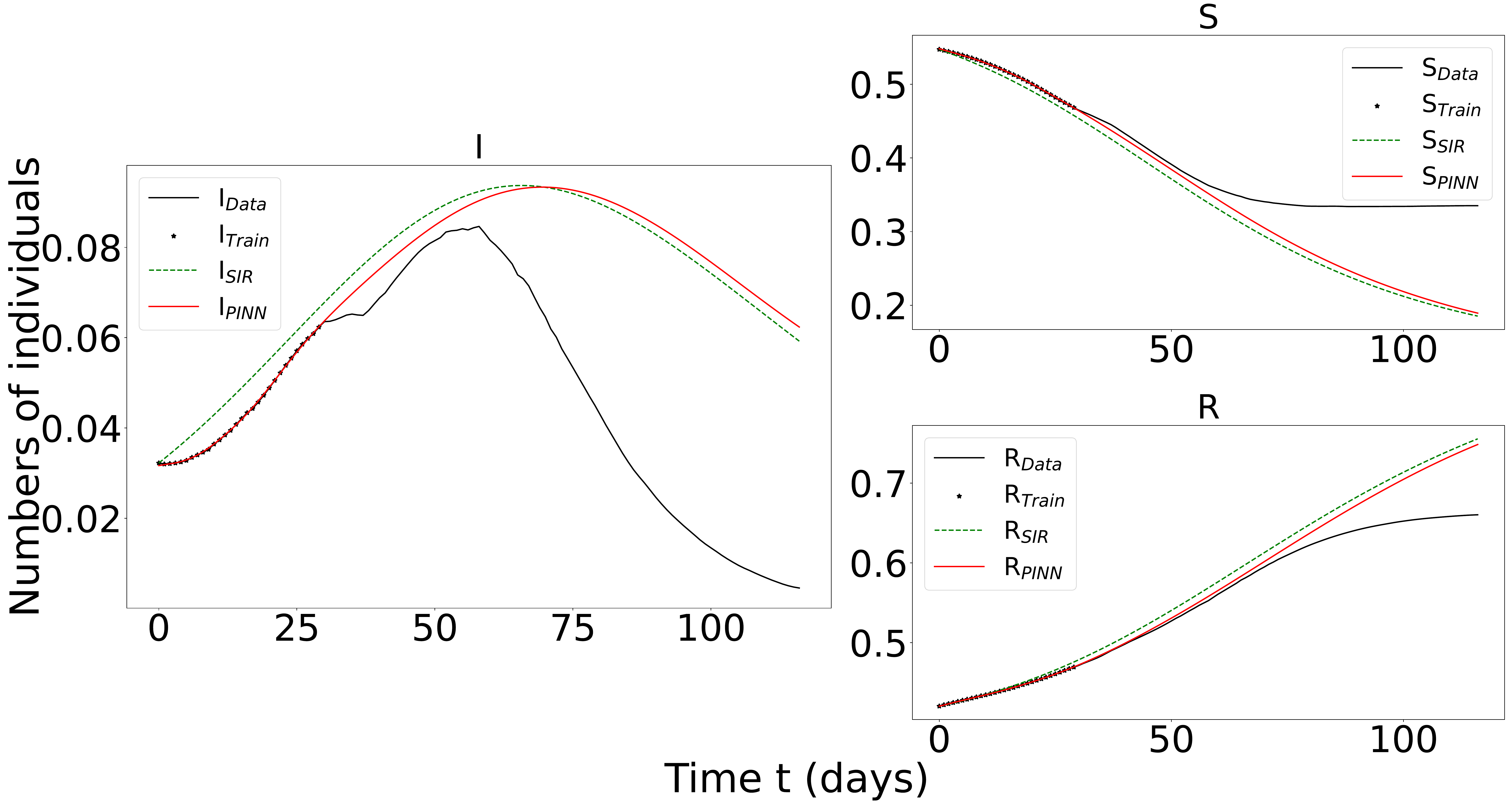}
 
    \caption{The graph shows the compartment results identified and predicted using the first 30 days as the training set. I, S and R diagrams of compartments I, S and R, respectively. The black star indicates the training set, the solid black line is the reported data, the red solid line is the result from PINNs, and the dashed green line indicates the result obtained by solving the SIR system using the parameters learned from PINNs.}
    \label{fig.realdata.30}

\end{figure}

\begin{figure}[ht]
	\centering
    \includegraphics[width=0.95\textwidth]{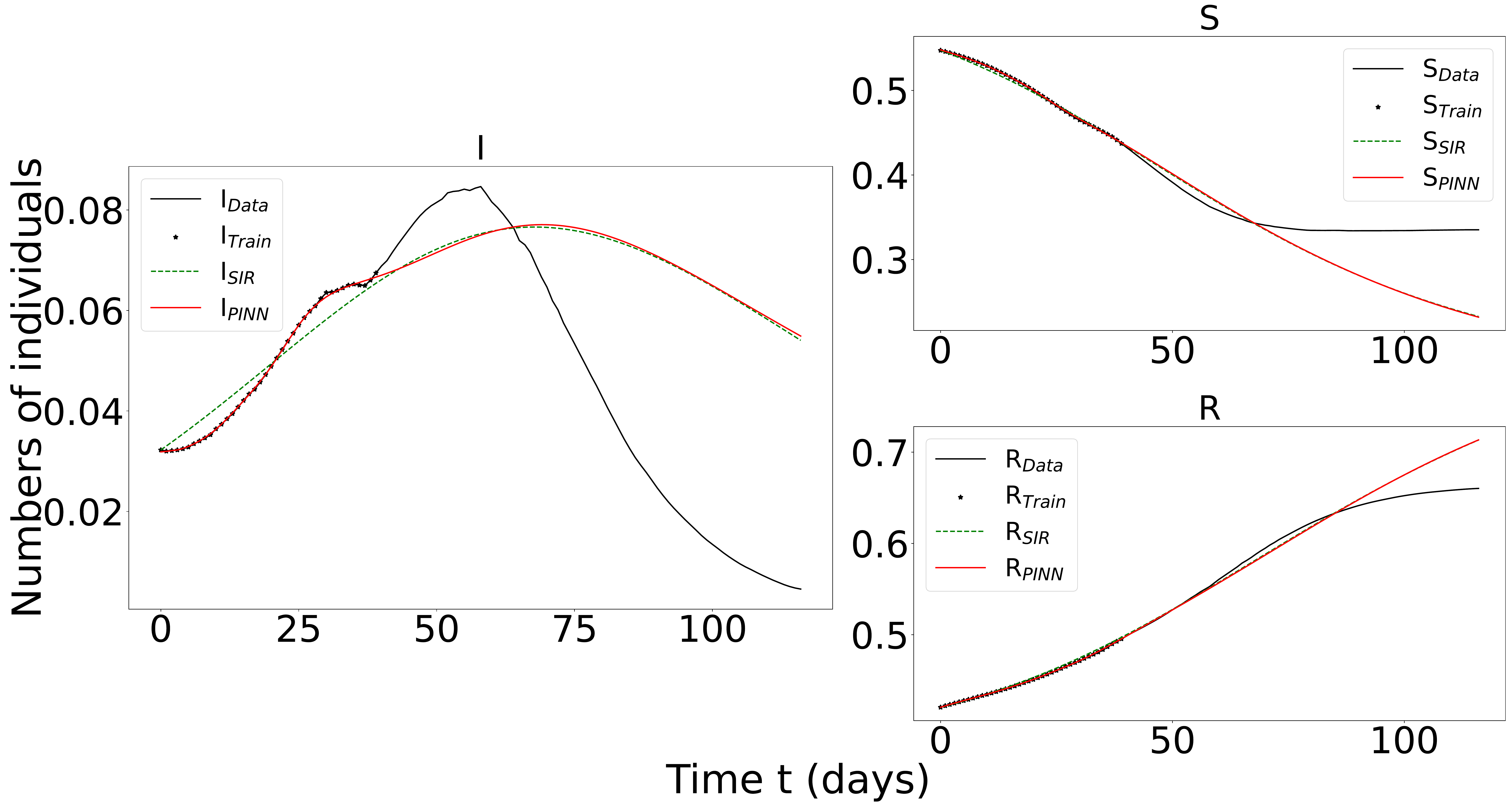}
 
    \caption{The graph shows the compartment results identified and predicted using the first 40 days as the training set. I, S and R diagrams of compartments I, S and R, respectively. The black star indicates the training set, the solid black line is the reported data, the red solid line is the result from PINNs, and the dashed green line indicates the result obtained by solving the SIR system using the parameters learned from PINNs.}
    \label{fig.realdata.40}

\end{figure}

\begin{figure}[ht]
	\centering
    \includegraphics[width=0.95\textwidth]{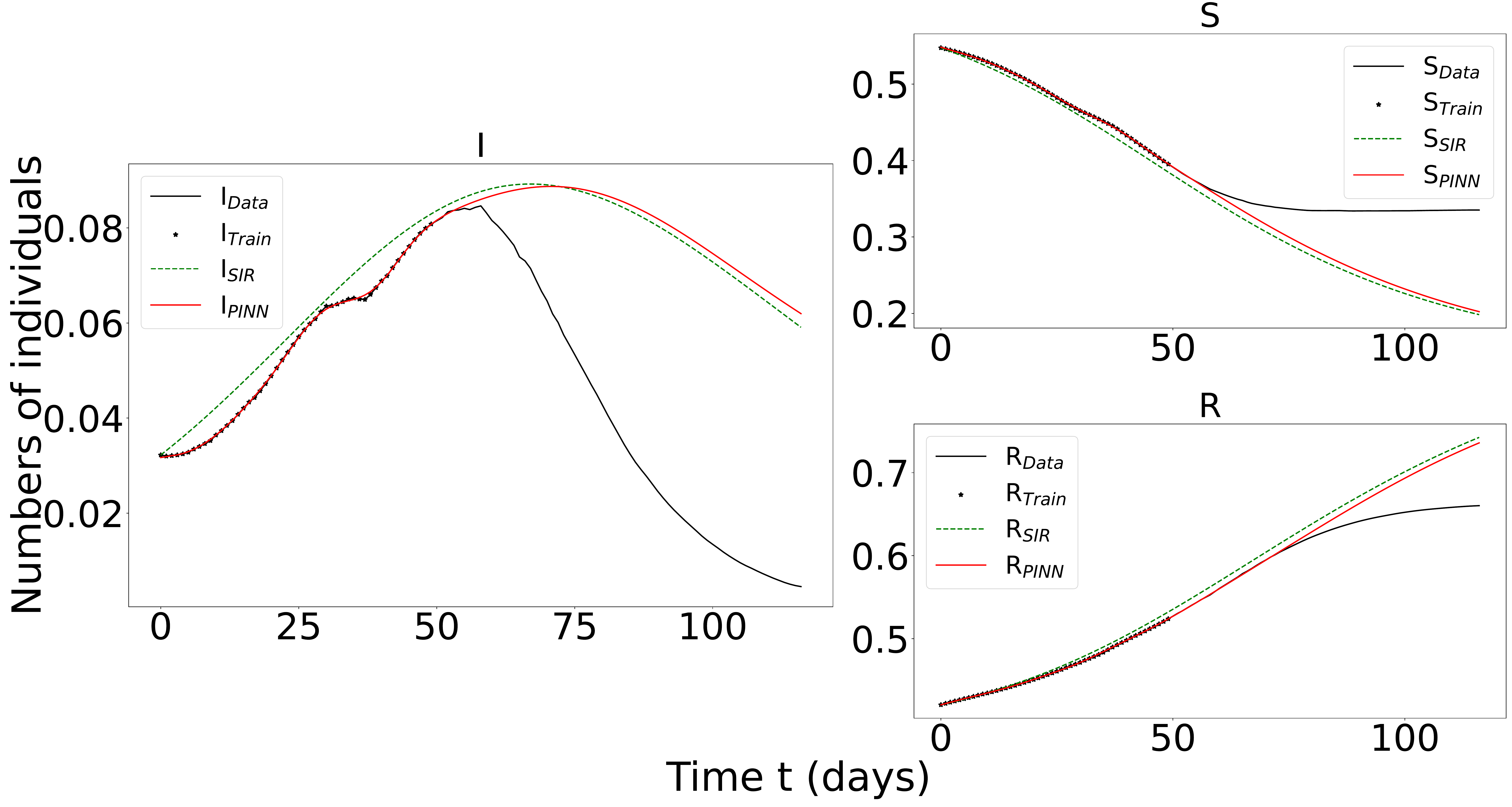}
 
    \caption{The graph shows the compartment results identified and predicted using the first 50 days as the training set. I, S and R diagrams of compartments I, S and R, respectively. The black star indicates the training set, the solid black line is the reported data, the red solid line is the result from PINNs, and the dashed green line indicates the result obtained by solving the SIR system using the parameters learned from PINNs.}
    \label{fig.realdata.50}

\end{figure}

After validation of the method on synthetic data, we move to test it on the realistic reported data. With pre-processing performed on the raw reported German data from March 1 to July 1, 2021, we selected the first 30, 40, and 50 days record of the compartment I (infected) as the training set and, to ensure reproductively, the same initial values of $\beta = 0.25$ and $\gamma = 0.15$ to the method for analysis and prediction testing. Figures~\ref{fig.realdata.30}, \ref{fig.realdata.40} and \ref{fig.realdata.50} show the identification and prediction results for using first 30, 40, 50 days as training set, respectively, including results from PINNs, solutions from the ODEs system \eqref{equ:sir} with PINN learned parameters, and the reported data (after pre-processing). In which, we adopt the LSODA algorithm \upcite{hindmarsh1983odepack} to solve the SIR model numerically with PINNs-learned parameters. 
%It is used to solve the numerical solution problem for rigid or non-rigid systems of first-order ordinary differential equations.

We can see that the proposed SIR-dynamics-informed neural network fits the training set well, and the trend of the predictions is also reasonably consistent with the true data especially at the peak appearance for I. Compared to the solutions of the ODE with parameters learned from the PINN, the results from SIR-informed NN have better a match on the training set e.g. on the I compartment while the prediction trend maintains, because of the balance introduced between the data loss and physical residual loss. In particular, the predictions of the proposed method are closer to the actual values around the peaks. It can be seen in the Table~\ref{tab:real_data_error}. Compared to purely data-driven machine learning method on the same problem, the residual loss contained in our method serves as a physics informed regularizer. On the other hand, the data-loss can also be viewed as a regularizer for the conventional physics dynamics fitting (e.g., using SIR or any other compartmental model). 
\par

Then, the PINNs learned parameters are used to produce the numerical solution of SIR model as Eq.\eqref{equ:sir} shows, which is obtained by using the LSODA algorithm \upcite{hindmarsh1983odepack} , which is used to solve the numerical solution problem for rigid or non-rigid systems of first-order ordinary differential equations. Then, the solution of the ODEs system is compared with the results from the PINNs.

%%%%%%%%%%%%%%%%%%%%%%%%%%%%%%%%%%%%%%%%%%%%%%%%%%%%%%%%%%%%%%%%%%%%%%%%%%%%%%%%%%%%%%%%%%
%%%%%%%%%%%%%%%%%%%%%%%%%%%%%%%%%%%%%%%%%%%%%%%%%%%%%%%%%%%%%%%%%%%%%%%%%%%%%%%%%%%%%%%%%%

%%%%%%%%%%%%%%%%%%%%%%%%%%%%%%%%%%%%%%%%%%%%%%%%%%%%%%%%%%%%%%%%%%%%%%%%%%%%%%%%%%%%%%%%%%%%
% \begin{figure}[hbtp!]
%     \begin{minipage}{.65\linewidth}
%         \centering
%         \subfloat[][I]{
%             \label{realdata.30:I}
%             \includegraphics[width=.95\linewidth]{figures/real_data/Germany_30_for_I_realdata.pdf}}
%     \end{minipage}
%     \medskip
%     \begin{minipage}{.35\linewidth}
%         \centering
%         \subfloat[][S]{
%             \label{realdata.30:S}
%             \includegraphics[width=.95\linewidth]{figures/real_data/Germany_30_for_S_realdata.pdf}}
    
%         \subfloat[][R]{
%             \label{realdata.30:R}
%             \includegraphics[width=.95\linewidth]{figures/real_data/Germany_30_for_R_realdata.pdf}}
%     \end{minipage} 
    
%     \caption{The graph shows the compartment results identified and predicted using the first 30 days as the training set. (a), (b) and (c) diagrams of compartments I, S and R, respectively. The black star indicates the training set, the solid black line is the reported data, the red solid line is the result from PINNs, and the dashed green line indicates the result obtained by solving the SIR system using the parameters learned from PINNs.}
%     \label{fig.realdata.30}
% \end{figure}
%%%%%%%%%%%%%%%%%%%%%%%%%%%%%%%%%%%%%%%%%%%%%%%%%%%%%%%%%%%%%%%%%%%%%%%%%%%%%%%%%%%%%%%%%

%%%%%%%%%%%%%%%%%%%%%%%%%%%%%%%%%%%%%%%%%%%%%%%%%%%%%%%%%%%%%%%%%%%%%%%%%%%%%%%%%%%%%%%%%%
%%%%%%%%%%%%%%%%%%%%%%%%%%%%%%%%%%%%%%%%%%%%%%%%%%%%%%%%%%%%%%%%%%%%%%%%%%%%%%%%%%%%%%%%%%

%%%%%%%%%%%%%%%%%%%%%%%%%%%%%%%%%%%%%%%%%%%%%%%%%%%%%%%%%%%%%%%%%%%%%%%%%%%%%%%%%%%%%%%%%%%%
% \begin{figure}[hbtp!]
%     \begin{minipage}{.65\linewidth}
%         \centering
%         \subfloat[][I]{
%             \label{realdata.40:I}
%             \includegraphics[width=.95\linewidth]{figures/real_data/Germany_40_for_I_realdata.pdf}}
%     \end{minipage}
%     \medskip
%     \begin{minipage}{.35\linewidth}
%         \centering
%         \subfloat[][S]{
%             \label{realdata.40:S}
%             \includegraphics[width=.95\linewidth]{figures/real_data/Germany_40_for_S_realdata.pdf}}
    
%         \subfloat[][R]{
%             \label{realdata.40:R}
%             \includegraphics[width=.95\linewidth]{figures/real_data/Germany_40_for_R_realdata.pdf}}
%     \end{minipage} 
    
%     \caption{The graph shows the compartment results identified and predicted using the first 40 days as the training set. (a), (b) and (c) diagrams of compartments I, S and R, respectively. The black star indicates the training set, the solid black line is the reported data, the red solid line is the result from PINNs, and the dashed green line indicates the result obtained by solving the SIR system using the parameters learned from PINNs.}
%     \label{fig.realdata.40}
% \end{figure}
%%%%%%%%%%%%%%%%%%%%%%%%%%%%%%%%%%%%%%%%%%%%%%%%%%%%%%%%%%%%%%%%%%%%%%%%%%%%%%%%%%%%%%%%%

%%%%%%%%%%%%%%%%%%%%%%%%%%%%%%%%%%%%%%%%%%%%%%%%%%%%%%%%%%%%%%%%%%%%%%%%%%%%%%%%%%%%%%%%%%
%%%%%%%%%%%%%%%%%%%%%%%%%%%%%%%%%%%%%%%%%%%%%%%%%%%%%%%%%%%%%%%%%%%%%%%%%%%%%%%%%%%%%%%%%%

%%%%%%%%%%%%%%%%%%%%%%%%%%%%%%%%%%%%%%%%%%%%%%%%%%%%%%%%%%%%%%%%%%%%%%%%%%%%%%%%%%%%%%%%%%%%
% \begin{figure}[hbtp!]
%     \begin{minipage}{.65\linewidth}
%         \centering
%         \subfloat[][I]{
%             \label{realdata.50:I}
%             \includegraphics[width=.95\linewidth]{figures/real_data/Germany_50_for_I_realdata.pdf}}
%     \end{minipage}
%     \medskip
%     \begin{minipage}{.35\linewidth}
%         \centering
%         \subfloat[][S]{
%             \label{realdata.50:S}
%             \includegraphics[width=.95\linewidth]{figures/real_data/Germany_50_for_S_realdata.pdf}}
    
%         \subfloat[][R]{
%             \label{realdata.50:R}
%             \includegraphics[width=.95\linewidth]{figures/real_data/Germany_50_for_R_realdata.pdf}}
%     \end{minipage} 
    
%     \caption{The graph shows the compartment results identified and predicted using the first 50 days as the training set. (a), (b) and (c) diagrams of compartments I, S and R, respectively. The black star indicates the training set, the solid black line is the reported data, the red solid line is the result from PINNs, and the dashed green line indicates the result obtained by solving the SIR system using the parameters learned from PINNs.}
%     \label{fig.realdata.50}
% \end{figure}
%%%%%%%%%%%%%%%%%%%%%%%%%%%%%%%%%%%%%%%%%%%%%%%%%%%%%%%%%%%%%%%%%%%%%%%%%%%%%%%%%%%%%%%%%
\par
%%%%%%%%%%%%%%%%%%%%%%%%%%%%%%%%%%%%%%%%%%%%%%%%%%%%%%%%%%%%%%%%%%%%%%%%%%%%%%%%%%%%%%%%%%
\begin{table}[hbtp!]
\centering
\caption{Error metrics for PINNs and SIR-ODEs on Germany reported data.}
\label{tab:real_data_error}
\begin{tabular}{@{}cccc@{}}
\toprule
    \multicolumn{2}{c}{}               & MSE        & MAE        \\ \midrule
    \multirow{2}*{30}      & PINNs     & 0.00644    & 0.08468    \\
                           & SIR-ODEs  & 0.00745    & 0.10278    \\ 
    \multirow{2}*{40}      & PINNs     & 0.00248    & 0.04984    \\
                           & SIR-ODEs  & 0.00244    & 0.05179    \\ 
    \multirow{2}*{50}      & PINNs     & 0.00499    & 0.06934    \\
                           & SIR-ODEs  & 0.00568    & 0.08474    \\ \midrule
\end{tabular}
\end{table}
%%%%%%%%%%%%%%%%%%%%%%%%%%%%%%%%%%%%%%%%%%%%%%%%%%%%%%%%%%%%%%%%%%%%%%%%%%%%%%%%%%%%%%%%%%%%%
\par
%%%%%%%%%%%%%%%%%%%%%%%%%%%%%%%%%%%%%%%%%%%%%%%%%%%%%%%%%%%%%%%%%%%%%%%%%%%%%%%%%%%%%%%%%%%%%
\begin{table}[hbtp!]
\centering
\caption{The result of parameter learned from PINNs on Germany reported data.}
\label{tab:parameters_earned_rea}
\begin{tabular}{@{}ccccc@{}}
\toprule
\multirow{4}*{} &  & \multicolumn{3}{c} {Parameters learned}            \\ \cmidrule(l){2-5} 
               & initial     & 30           & 40           & 50         \\ \midrule
$\beta$        & 0.25        & 0.1268       & 0.1167       & 0.1233     \\
$\gamma$       & 0.15        & 0.0392       & 0.0404       & 0.0391     \\ \midrule  

\end{tabular}
\end{table}
%%%%%%%%%%%%%%%%%%%%%%%%%%%%%%%%%%%%%%%%%%%%%%%%%%%%%%%%%%%%%%%%%%%%%%%%%%%%%%%%%%%%%%%%%%
\par
We found that in the process of identifying dynamics and makeing forecasting on reported German data using PINN, several factors have an impact on the training convergence speed and final results: the initial value setting of the trainable parameters, the different assignment of the data loss, and residual loss weights in the loss function. However, this effect is negligible for synthetic data, which is less noisy and with cleaner dynamics. In Part A of the Appendix, we analyzed the impact of different weight assignments for the loss function's data loss and residual loss on the error and the prediction and identification of the number of infections. The goal is to analyze which weighting assignment of the data and residual loss in the loss function can be optimal for the error impact of the results, i.e., lead to the lowest error and more accurate identification and prediction in compartment I.

\section{Conclusions}
\label{sec: conclusions}
A machine learning approach based on physics-informed neural networks is considered to identify, predict and estimate, the parameters for the evolution of dynamic systems. We combine it with a simple SIR compartmental model from mathematical epidemiology to learn and explore complex epidemiological dynamics. We first demonstrate the proposed method on synthetic data from a SAIRD model in multiple numerical experiments, with different-sized training sets generated by solving a system of ODEs using fixed parameters and initial conditions. It's found that the PINNs with SIR model performs better on the training set and provided a better trend of the prediction than solutions from ODEs just with model parameters estimated using PINNs. \par

Then, we extended our tests to realistic reported data about the COVID-19 epidemic in Germany. Specifically we picked a period in which compartment I had a peak for demonstration. The experimental results show that the PINNs combined with the SIR model also gives better results in the training set and trend prediction for the test set than solving the ODE system using the parameters estimated by the PINNs. Experimental results from both simulations and reported data show that it is feasible to put physical information from simple epidemic models into PINNs to study more complex epidemiological dynamics. Moreover, with certain finetuning, better results can be achieved. \par

However, in order to obtain more accurate results in actual epidemiological data, we should extend our study to more detailed models in mathematical epidemiology while explore other architectures of PINNs. This will be the subject of future work.\par

\section*{Funding}
The present work was supported by XF-IJRC (S. Han, L. Wang), the ENABLE Project of HMWK (L. Stelz), the BMBF under ErUM-Data (K. Zhou), the AI grant of SAMSON AG, Frankfurt (L. Wang, K. Zhou), the Walter Greiner Gesellschaft zur F\"orderung der physikalischen Grundlagenforschung e.V. through the Judah M. Eisenberg Laureatus Chair at Goethe Universit\"at Frankfurt am Main (H. St\"ocker). We also thank NVIDIA Corporation for the donation of NVIDIA GPUs.

%\section*{Acknowledgments}
%The authors thank XXX .

\section*{Data Availability Statement}
The reported COVID-19 data for Germany used for the present work are available at the COVID-19 Data Repository (https://github.com/robert-koch-institut) and provided by the German Robert Koch Institute (RKI Berlin, Germany) data.

% The Germany COVID-19 reported data used during the current study are available at COVID-19 Data Repository (https://github.com/robert-koch-institut), from the Robert Koch Institute (RKI) Berlin, Germany data.

% The Germany COVID-19 reported data used during the current study are available at COVID-19 Data Repository by the Center for Systems Science and Engineering (CSSE) at Johns Hopkins University (https://github.com/CSSEGISandData/COVID-19) which is derived from the German Robert Koch Institute (RKI) statistics.

\section*{Conflicts of Interest}
The authors declare that they do not see any conflict of interest.

\bibliographystyle{elsarticle-num} 
\bibliography{cas-refs}

\begin{thebibliography}{10}
\expandafter\ifx\csname url\endcsname\relax
  \def\url#1{\texttt{#1}}\fi
\expandafter\ifx\csname urlprefix\endcsname\relax\def\urlprefix{URL }\fi
\expandafter\ifx\csname href\endcsname\relax
  \def\href#1#2{#2} \def\path#1{#1}\fi

\bibitem{WHOcoronavirus2022}
WHO, \href{https://covid19.who.int/}{Who coronavirus for covid-19 on 2022.},
  World Health Organization (2022).
\newline\urlprefix\url{https://covid19.who.int/}

\bibitem{sedaghat2020predicting}
A.~Sedaghat, S.~Band, A.~Mosavi, L.~Nadai, Predicting trends of coronavirus
  disease (covid-19) using sird and gaussian-sird models, in: 2020 IEEE 3rd
  International Conference and Workshop in {\'O}buda on Electrical and Power
  Engineering (CANDO-EPE), IEEE, 2020, pp. 000267--000274.

\bibitem{siegenfeld2020models}
A.~F. Siegenfeld, N.~N. Taleb, Y.~Bar-Yam, What models can and cannot tell us
  about covid-19, Proceedings of the National Academy of Sciences 117~(28)
  (2020) 16092--16095.

\bibitem{chinazzi2020effect}
M.~Chinazzi, J.~T. Davis, M.~Ajelli, C.~Gioannini, M.~Litvinova, S.~Merler,
  A.~Pastore~y Piontti, K.~Mu, L.~Rossi, K.~Sun, et~al., The effect of travel
  restrictions on the spread of the 2019 novel coronavirus (covid-19) outbreak,
  Science 368~(6489) (2020) 395--400.

\bibitem{wilder2020isolation}
A.~Wilder-Smith, D.~O. Freedman, Isolation, quarantine, social distancing and
  community containment: pivotal role for old-style public health measures in
  the novel coronavirus (2019-ncov) outbreak., Journal of travel medicine
  (2020).

\bibitem{zhang2020covid}
S.~Zhang, Z.~Wang, R.~Chang, H.~Wang, C.~Xu, X.~Yu, L.~Tsamlag, Y.~Dong,
  H.~Wang, Y.~Cai, Covid-19 containment: China provides important lessons for
  global response, Frontiers of Medicine 14~(2) (2020) 215--219.

\bibitem{banholzer2021estimating}
N.~Banholzer, E.~Van~Weenen, A.~Lison, A.~Cenedese, A.~Seeliger, B.~Kratzwald,
  D.~Tschernutter, J.~P. Salles, P.~Bottrighi, S.~Lehtinen, et~al., Estimating
  the effects of non-pharmaceutical interventions on the number of new
  infections with covid-19 during the first epidemic wave, PLoS one 16~(6)
  (2021) e0252827.

\bibitem{thiel2021recommendations}
N.~Thiel, C.~Selwyn, G.~Murphy, S.~Simpson, A.~C. Chakrabarti, Recommendations
  for acceleration of vaccine development and emergency use filings for
  covid-19 leveraging lessons from the novel oral polio vaccine, npj Vaccines
  6~(1) (2021) 1--8.

\bibitem{cao2022covid}
L.~Cao, Q.~Liu, Covid-19 modeling: A review, medRxiv (2022).

\bibitem{park2020time}
S.~W. Park, D.~M. Cornforth, J.~Dushoff, J.~S. Weitz, The time scale of
  asymptomatic transmission affects estimates of epidemic potential in the
  covid-19 outbreak, Epidemics 31 (2020) 100392.

\bibitem{dorn2023common}
F.~Dorn, S.~Khailaie, M.~Stoeckli, S.~C. Binder, T.~Mitra, B.~Lange,
  S.~Lautenbacher, A.~Peichl, P.~Vanella, T.~Wollmersh{\"a}user, et~al., The
  common interests of health protection and the economy: evidence from scenario
  calculations of covid-19 containment policies, The European Journal of Health
  Economics 24~(1) (2023) 67--74.

\bibitem{TANWAR2022108352}
S.~Tanwar, A.~Kumari, D.~Vekaria, N.~Kumar, R.~Sharma, An ai-based disease
  detection and prevention scheme for covid-19, Computers and Electrical
  Engineering 103 (2022) 108352.

\bibitem{ertas2021role}
Y.~N. Ertas, M.~Mahmoodi, F.~Shahabipour, V.~Jahed, S.~E. Diltemiz, R.~Tutar,
  N.~Ashammakhi, Role of biomaterials in the diagnosis, prevention, treatment,
  and study of corona virus disease 2019 (covid-19), Emergent materials 4~(1)
  (2021) 35--55.

\bibitem{kumar2020does}
S.~Kumar, S.~Managi, Does stringency of lockdown affect air quality? evidence
  from indian cities, Economics of Disasters and Climate Change 4~(3) (2020)
  481--502.

\bibitem{papadopoulos2020impact}
D.~I. Papadopoulos, I.~Donkov, K.~Charitopoulos, S.~Bishara, The impact of
  lockdown measures on covid-19: a worldwide comparison, MedRxiv (2020).

\bibitem{barbarossa2020first}
M.~V. Barbarossa, J.~Fuhrmann, J.~Heidecke, H.~V. Varma, N.~Castelletti, J.~H.
  Meinke, S.~Krieg, T.~Lippert, A first study on the impact of current and
  future control measures on the spread of covid-19 in germany, medRxiv (2020)
  2020--04.

\bibitem{barbarossa2021germany}
M.~V. Barbarossa, J.~Fuhrmann, Germany’s next shutdown—possible scenarios
  and outcomes, Influenza and other respiratory viruses 15~(3) (2021) 326--330.

\bibitem{peeri2020sars}
N.~C. Peeri, N.~Shrestha, M.~S. Rahman, R.~Zaki, Z.~Tan, S.~Bibi,
  M.~Baghbanzadeh, N.~Aghamohammadi, W.~Zhang, U.~Haque, The sars, mers and
  novel coronavirus (covid-19) epidemics, the newest and biggest global health
  threats: what lessons have we learned?, International journal of epidemiology
  49~(3) (2020) 717--726.

\bibitem{xiang2021covid}
Y.~Xiang, Y.~Jia, L.~Chen, L.~Guo, B.~Shu, E.~Long, Covid-19 epidemic
  prediction and the impact of public health interventions: A review of
  covid-19 epidemic models, Infectious Disease Modelling 6 (2021) 324--342.

\bibitem{cooper2020dynamic}
I.~Cooper, A.~Mondal, C.~G. Antonopoulos, Dynamic tracking with model-based
  forecasting for the spread of the covid-19 pandemic, Chaos, Solitons \&
  Fractals 139 (2020) 110298.

\bibitem{kermack1927contribution}
W.~O. Kermack, A.~G. McKendrick, A contribution to the mathematical theory of
  epidemics, Proceedings of the royal society of london. Series A, Containing
  papers of a mathematical and physical character 115~(772) (1927) 700--721.

\bibitem{aron1984seasonality}
J.~L. Aron, I.~B. Schwartz, Seasonality and period-doubling bifurcations in an
  epidemic model, Journal of theoretical biology 110~(4) (1984) 665--679.

\bibitem{sun2020seir}
P.~Sun, K.~Li, An seir model for assessment of current covid-19 pandemic
  situation in the uk, medRxiv (2020).

\bibitem{basnarkov2021seair}
L.~Basnarkov, Seair epidemic spreading model of covid-19, Chaos, Solitons \&
  Fractals 142 (2021) 110394.

\bibitem{dandekar2020quantifying}
R.~Dandekar, G.~Barbastathis, Quantifying the effect of quarantine control in
  covid-19 infectious spread using machine learning, MedRxiv (2020).

\bibitem{jamshidi2020artificial}
M.~Jamshidi, A.~Lalbakhsh, J.~Talla, Z.~Peroutka, F.~Hadjilooei, P.~Lalbakhsh,
  M.~Jamshidi, L.~La~Spada, M.~Mirmozafari, M.~Dehghani, et~al., Artificial
  intelligence and covid-19: deep learning approaches for diagnosis and
  treatment, Ieee Access 8 (2020) 109581--109595.

\bibitem{vaid2020deep}
S.~Vaid, R.~Kalantar, M.~Bhandari, Deep learning covid-19 detection bias:
  accuracy through artificial intelligence, International Orthopaedics 44~(8)
  (2020) 1539--1542.

\bibitem{stout2021silent}
R.~L. Stout, S.~J. Rigatti, The silent pandemic covid-19 in the asymptomatic
  population, medRxiv (2021) 2020--12.

\bibitem{worby2020face}
C.~J. Worby, H.-H. Chang, Face mask use in the general population and optimal
  resource allocation during the covid-19 pandemic, Nature communications
  11~(1) (2020) 1--9.

\bibitem{oraby2021modeling}
T.~Oraby, M.~G. Tyshenko, J.~C. Maldonado, K.~Vatcheva, S.~Elsaadany, W.~Q.
  Alali, J.~C. Longenecker, M.~Al-Zoughool, Modeling the effect of lockdown
  timing as a covid-19 control measure in countries with differing social
  contacts, Scientific reports 11~(1) (2021) 1--13.

\bibitem{choi2021optimal}
W.~Choi, E.~Shim, Optimal strategies for social distancing and testing to
  control covid-19, Journal of theoretical biology 512 (2021) 110568.

\bibitem{barbarossa2020modeling}
M.~V. Barbarossa, J.~Fuhrmann, J.~H. Meinke, S.~Krieg, H.~V. Varma,
  N.~Castelletti, T.~Lippert, Modeling the spread of covid-19 in germany: Early
  assessment and possible scenarios, Plos one 15~(9) (2020) e0238559.

\bibitem{hassan2020covid}
A.~Hassan, I.~Shahin, M.~B. Alsabek, Covid-19 detection system using recurrent
  neural networks, in: 2020 International conference on communications,
  computing, cybersecurity, and informatics (CCCI), IEEE, 2020, pp. 1--5.

\bibitem{wang2021machine}
L.~Wang, T.~Xu, T.~Stoecker, H.~Stoecker, Y.~Jiang, K.~Zhou, Machine learning
  spatio-temporal epidemiological model to evaluate germany-county-level
  covid-19 risk, Machine Learning: Science and Technology 2~(3) (2021) 035031.

\bibitem{chimmula2020time}
V.~K.~R. Chimmula, L.~Zhang, Time series forecasting of covid-19 transmission
  in canada using lstm networks, Chaos, Solitons \& Fractals 135 (2020) 109864.

\bibitem{zeroual2020deep}
A.~Zeroual, F.~Harrou, A.~Dairi, Y.~Sun, Deep learning methods for forecasting
  covid-19 time-series data: A comparative study, Chaos, Solitons \& Fractals
  140 (2020) 110121.

\bibitem{qin2019data}
T.~Qin, K.~Wu, D.~Xiu, Data driven governing equations approximation using deep
  neural networks, Journal of Computational Physics 395 (2019) 620--635.

\bibitem{chen2021generalized}
Z.~Chen, D.~Xiu, On generalized residual network for deep learning of unknown
  dynamical systems, Journal of Computational Physics 438 (2021) 110362.

\bibitem{nguyen2020artificial}
T.~T. Nguyen, Q.~V.~H. Nguyen, D.~T. Nguyen, E.~B. Hsu, S.~Yang, P.~Eklund,
  Artificial intelligence in the battle against coronavirus (covid-19): a
  survey and future research directions, arXiv preprint arXiv:2008.07343
  (2020).

\bibitem{chen2021survey}
J.~Chen, K.~Li, Z.~Zhang, K.~Li, P.~S. Yu, A survey on applications of
  artificial intelligence in fighting against covid-19, ACM Computing Surveys
  (CSUR) 54~(8) (2021) 1--32.

\bibitem{ganslmeier2021impact}
M.~Ganslmeier, D.~Furceri, J.~D. Ostry, The impact of weather on covid-19
  pandemic, Scientific reports 11~(1) (2021) 1--7.

\bibitem{raissi2019physics}
M.~Raissi, P.~Perdikaris, G.~E. Karniadakis, Physics-informed neural networks:
  A deep learning framework for solving forward and inverse problems involving
  nonlinear partial differential equations, Journal of Computational physics
  378 (2019) 686--707.

\bibitem{wang2021physics}
R.~Wang, R.~Yu, Physics-guided deep learning for dynamical systems: A survey,
  arXiv preprint arXiv:2107.01272 (2021).

\bibitem{karniadakis2021physics}
G.~E. Karniadakis, I.~G. Kevrekidis, L.~Lu, P.~Perdikaris, S.~Wang, L.~Yang,
  Physics-informed machine learning, Nature Reviews Physics 3~(6) (2021)
  422--440.

\bibitem{kharazmi2021identifiability}
E.~Kharazmi, M.~Cai, X.~Zheng, Z.~Zhang, G.~Lin, G.~E. Karniadakis,
  Identifiability and predictability of integer-and fractional-order
  epidemiological models using physics-informed neural networks, Nature
  Computational Science 1~(11) (2021) 744--753.

\bibitem{angeli2022modeling}
M.~Angeli, G.~Neofotistos, M.~Mattheakis, E.~Kaxiras, Modeling the effect of
  the vaccination campaign on the covid-19 pandemic, Chaos, Solitons \&
  Fractals 154 (2022) 111621.

\bibitem{Kingma2014adam}
D.~P. Kingma, J.~Ba, Adam: A method for stochastic optimization (2014).
\newblock \href {https://doi.org/10.48550/ARXIV.1412.6980}
  {\path{doi:10.48550/ARXIV.1412.6980}}.

\bibitem{hindmarsh1983odepack}
A.~C. Hindmarsh, Odepack: A systemized collection of ode solvers, Scientific
  computing (1983) 55--64.

\end{thebibliography}

\newpage

\section*{Appendix A: Sensitivity analysis for weight setup}
The PINN's loss function consists of two parts, the data loss and the residual loss.
%%%%%%%%%%%%%%%%%%%%%%%%%%%%%%%%%%%%%%%%%%%%%%%%%%%%%%%%%%%%%%%%%%%%%%%%%%%%%%%%%%%%%%%%%%%
\begin{equation}
    \begin{aligned}
        LOSS = \alpha_1 MSE_{data} + \alpha_2 MSE_{residual}.
    \end{aligned}
\end{equation}
%%%%%%%%%%%%%%%%%%%%%%%%%%%%%%%%%%%%%%%%%%%%%%%%%%%%%%%%%%%%%%%%%%%%%%%%%%%%%%%%%%%%%%%%%%%
Here $\alpha_1$ and $\alpha_2$ are the weights for data loss and residual loss, respectively. 
%%%%%%%%%%%%%%%%%%%%%%%%%%%%%%%%%%%%%%%%%%%%%%%%%%%%%%%%%%%%%%%%%%%%%%%%%%%%%%%%%%%%%%%%%%%
% \usepackage{booktabs}
\begin{table}[ht!]
% \footnotesize
\centering
\caption{The different allocations of data loss and residual loss are compared with the Mean Average Error(MAE) and the Mean Squared Error(MSE) of the respective results. Here $\alpha_1$ is the weight of the data loss and $\alpha_2$ is the weight of the residual loss.}
\label{tab: sensitive analysis}
\begin{tabular}{@{}llllllll@{}}
\toprule
$(\alpha_1, \alpha_2)$ & (1,1) & (5,3) & (10,5) & (15,7) & (20,9) & (25,11) & (30,13) \\ \midrule
MAE                    &0.04240       &0.03557       &0.03551        &0.03476        &0.03444        &0.03376          &0.03423          \\
MSE                    &0.00145       &0.00107       &0.00103        &0.00098        &0.00097        &0.00092         &0.00095         \\ \bottomrule

% $\alpha_1$ & $\alpha_2$ & MAE & MSE \\ \midrule
% 1        & 1         & 0.04240    & 0.00145  \\
% 5        & 3         & 0.03557    & 0.00107   \\
% 10       & 5         & 0.03551    & 0.00103  \\
% 15       & 7         & 0.03476    & 0.00098  \\
% 20       & 9         & 0.03444    & 0.00097  \\
% 25       & 11        & 0.03376    & 0.00092  \\
% 30       & 13        & 0.03423    & 0.00095  \\ \bottomrule
\end{tabular}
\end{table}
\end{sloppypar}
\end{document}